\documentclass[aps,pra,twocolumn,a4paper,showpacs]{revtex4}
\usepackage{amsmath}
\usepackage{amssymb}
\usepackage{amsfonts}
\usepackage{txfonts}
\usepackage{bbold}
\usepackage{color}
\usepackage{bm}
\usepackage{graphicx}
\begin{document}

\title{The manifestation of quantum resonances and antiresonances \\ in a finite temperature dilute atomic gas}

\author{M. Saunders}
\affiliation{Department of Physics, Durham University,
Rochester Building, South Road, Durham DH1 3LE, United Kingdom}

\author{P. L. Halkyard}
\affiliation{Department of Physics, Durham University,
Rochester Building, South Road, Durham DH1 3LE, United Kingdom}

\author{K. J. Challis}
\affiliation{Department of Physics, Durham University,
Rochester Building, South Road, Durham DH1 3LE, United Kingdom}

\author{S. A. Gardiner}
\affiliation{Department of Physics, Durham University,
Rochester Building, South Road, Durham DH1 3LE, United Kingdom}

\date{\today}

\begin{abstract}

We investigate the effect of temperature on resonant and antiresonant dynamics in a dilute atomic gas kicked periodically by a standing wave laser field.  Our numerical calculations are based on a Monte Carlo method for an incoherent mixture of non-interacting plane waves, and show that the atomic dynamics are highly sensitive to the initial momentum width of the gas.  We explain this sensitivity by examining the time evolution of individual atomic centre of mass momentum eigenstates with varying quasimomentum, and we have determined analytic expressions for the evolution of the second-order momentum moment to illustrate the range of behaviours.
\end{abstract}


\maketitle

\section{Introduction \label{Sec:Introduction}}

A dilute atomic gas subjected to a pulsed optical standing wave is, under certain conditions, an accurate realization of a periodically kicked quantum rotor~\cite{Moore1995,Ammann1998a,dArcy2001b}.  The  kicked rotor is a physical realization of the standard map, which is a well established model system in the study of classically chaotic Hamiltonian dynamics~\cite{Lichtenberg1992,Reichl2004}.  The analogous quantum kicked rotor provides the opportunity to probe quantum dynamics in regimes where the underlying classical system is chaotic \cite{Reichl2004,Haake2001}.  

The atom-optical realization of the kicked rotor, where laser-cooled atoms are periodically driven by a far-detuned laser, is a very attractive system to study because of the precise control available in experiments.  The amplitude, phase, and length of the kicks can be varied over a wide parameter regime, and the initial state of the gas can be carefully chosen using the sophisticated cooling and confinement methods now available \cite{Phillips1998}.  A large number of such laser-driven cold atom experiments have been carried out \cite{Moore1995,Ammann1998a,dArcy2001b,Oskay2000,dArcy2001a,Sadgrove2004,Bharucha1999,Moore1994,Klappauf1999,Steck2000,Milner2000,Oskay2003,Vant2000,Doherty2000,Kanem2007,Oberthaler1999,Godun2000,Schlunk2003a,Schlunk2003b,Ma2004,Buchleitner2006,dArcy2003,Duffy2004b,Behinaein2006, Ryu2006,Szriftgiser2002,Ammann1998b,Vant1999,Williams2004,Duffy2004a}, and these have led to the observation of quantum resonant dynamics~\cite{Bharucha1999,Oskay2000, dArcy2001a,Sadgrove2004}, dynamical localisation~\cite{Moore1994,Moore1995,Klappauf1999,Bharucha1999,Steck2000,Milner2000,Oskay2003,Ammann1998a,Vant2000,Doherty2000,dArcy2001b}, higher-order resonances~\cite{Kanem2007},  and quantum accelerator modes~\cite{Oberthaler1999,Godun2000,dArcy2001a,Schlunk2003a,Schlunk2003b,Ma2004,Buchleitner2006}.  In the ultra-cold regime, a Bose-Einstein condensed cloud has made it possible to observe quantum antiresonances~\cite{Duffy2004b} and provide detailed information about the momentum distribution of a $\delta$-kicked gas~\cite{Behinaein2006, Ryu2006}.

Such atom-optical experimental configurations are clearly not exact realizations of the ideal quantum $\delta$-kicked rotor.  For example, absolutely instantaneous kicks are impossible to achieve, and interatomic collisions may also come into play.  A more fundamental difference is the fact that a laser-driven atom models a $\delta$-kicked free particle, rather than a rotor.  The angular momentum spectrum of a quantum rotor is discrete, whereas the momentum spectrum of a free particle is continuous.  As first noted by Bharucha \textit{et al.}~\cite{Bharucha1999}, this can substantially change the manifestation of quantum resonant and antiresonant dynamics.

In this paper we investigate the quantum-resonant and antiresonant behaviour of clouds of $\delta$-kicked cold atoms \cite{Dana1995,Dana1996,Dana2005,Dana2006a,Dana2006b,Wimberger2003,Wimberger2004,Wimberger2005a,Wimberger2005b} as the width of the initial centre of mass atomic momentum distribution is varied.  In particular, we investigate the transition from the zero-temperature limit, where the expected behaviour of a $\delta$-kicked rotor is exactly reproduced, to a qualitatively different behaviour typical of atomic samples initially laser-cooled to microKelvin temperatures.

The paper is organised as follows.  In Sec.~\ref{sec:background} we outline our theoretical model for the atom-optical kicked rotor.  We begin by deriving our model Hamiltonian for the system in Sec.~\ref{sec:model}, and give an analytic description of the temporal dynamics in Sec.~\ref{sec:time}.  In Sec.~\ref{sec:numerical} we outline our numerical implementation of the theory, and reproduce some known results to illustrate the method.  In Sec.~\ref{sec:momentum} we consider the dynamics of a periodically kicked plane wave and present a detailed survey of the effect of the initial momentum on the system behaviour.  We derive analytic expressions for the positions and widths of the resonances and antiresonances observed, and identify a number of higher-order antiresonance features.  In Sec.~\ref{sec:temperature} we model periodically kicked atom clouds with a finite initial momentum width.  We consider the temperature dependence of the system dynamics, and explain the results based on our understanding of resonances and antiresonances in the plane-wave case.

\section{Background \label{sec:background}}

\subsection{Model Hamiltonian \label{sec:model}}

\subsubsection{Two-level atom in a laser field}

We consider a cloud of trapped and laser-cooled thermal alkali atoms.  The cloud is released from all external fields, and then addressed by two far-detuned laser beams that form a pulsed sinusoidal potential aligned perpendicular to gravity~\footnote{In a vertical laser configuration, the gravitational acceleration can be compensated for~ \cite{dArcy2001a, Ma2004}, e.g., by controlling the relative phase difference between the two laser fields~\cite{Denschlag2002}.}.  We neglect atomic collisions in the gas, allowing for a theoretical description using a single-particle Hamiltonian.  This is reasonable for a dilute thermal gas.  When considering experiments carried out on an initially Bose-Einstein condensed sample~\cite{Duffy2004b, Behinaein2006, Ryu2006, Sadgrove2007, Ramareddy2007}, a noninteracting model serves as a useful starting point.  Furthermore, near a Feshbach resonance, magnetic fields can be used to tune the scattering length, in principle to zero (e.g., \cite{Inouye1998,Roberts1998,Kohler2006,Kraemer2004}).  The cloud dynamics are described using a one-dimensional treatment along the axis of the standing wave.  In our noninteracting model, this approach applies without loss of generality because the equation of motion for the three spatial degrees of freedom is separable.

Our model Hamiltonian $\hat{H}$ describes a single two-level atom of mass~$M$, with internal ground state $|g\rangle$ and excited state $|e\rangle$.  The internal atomic levels are coupled by two equal-frequency laser travelling waves with a controllable phase difference \cite{Denschlag2002}.  As we always consider the lasers to be far-detuned (see Sec.~\ref{adiabatic}), we neglect spontaneous emission from the outset. Hence,
\begin{equation}
\begin{split}
\hat{H} = &
\frac{1}{2}\hbar\omega_{0} \left( |e\rangle\langle e|-|g\rangle\langle g| \right)
+
\frac{\hat{p}^{2}}{2M}
\\&
+
\frac{1}{2}\hbar\Omega_{1}
\left[
e^{i(k_{L}\hat{x}-\omega_{L}t+\phi_{1})}
|e\rangle \langle g|
+ \mbox{H.c.}
\right]
\\&
+
\frac{1}{2}\hbar\Omega_{2}
\left[
e^{-i(k_{L}\hat{x}+\omega_{L}t-\phi_{2})}
|e\rangle \langle g|
+ \mbox{H.c.}
\right],
\end{split}
\label{Hamiltonian_all}
\end{equation}
where $\hat{x}$ is the one-dimensional centre-of-mass atomic position operator with $\hat{p}$ its conjugate momentum, $t$ is the time, $\hbar\omega_{0}$ is the energy difference between the internal atomic levels, $\Omega_{1}$ and $\Omega_{2}$ are the laser Rabi frequencies with $\phi_{1}$ and $\phi_{2}$ the respective laser phases, $\omega_{L}$ is the laser frequency, and $k_{L}$~is the laser wave-vector magnitude along the $x$ direction.

Assuming that the two laser fields are of equal intensity and phase (i.e., $\Omega_{1}=\Omega_{2}=\Omega/2$ and $\phi_{1}=\phi_{2} = \phi$),
\begin{equation}
\begin{split}
\hat{H}  = &
\frac{1}{2}\hbar\omega_{0}(|e\rangle\langle e|-|g\rangle\langle g|)
+
\frac{\hat{p}^{2}}{2M}
\\
&+\frac{1}{2}\hbar\Omega
\cos(k_{L}\hat{x})
\left[
e^{-i(\omega_{L}t-\phi)}|e \rangle\langle g|
+ \mbox{H.c.}
\right].
\end{split}
\label{H_rearrange}
\end{equation}
To eliminate the time-dependence in Eq.~(\ref{H_rearrange}), and shift the energy zero to the atomic ground state, we carry out a unitary transformation defined by
$
\hat{U}_1 = \exp(i[\omega_{L}|e\rangle\langle e| -\omega_{0}(|e\rangle \langle e| + |g\rangle\langle g|)/2]t)
$.  Hamiltonian~(\ref{H_rearrange}) then transforms to
\begin{equation}
\hat{H}' = \hbar\Delta |e\rangle \langle e| +
\frac{\hat{p}^{2}}{2M}
+
\frac{1}{2}\hbar \Omega\cos(k_{L}\hat{x})
\left[
e^{i\phi}|e\rangle \langle g | + \mbox{H.c.}
\right],
\end{equation}
where we have defined the detuning $\Delta \equiv \omega_{0}-\omega_{L}$.

\subsubsection{Adiabatic elimination of the excited state \label{adiabatic}}

In the far-detuned limit ($\Omega/\Delta \ll 1$), and for the case that all the
atomic population is initially in the internal ground state~$|g\rangle$, the
excited state population is always negligible.  In this regime, the excited
state~$|e\rangle$ can be adiabatically eliminated~\cite{Meystre1991}.  If we
also invoke the unitary transformation $\hat{U}_2=\exp(-i\Omega^2|g \rangle \langle g|t/8\Delta)$, the Hamiltonian simplifies to 
\begin{equation}
\hat{H}'' =\frac{ \hat{p}^{2}}{2M}  - \frac{\hbar \Omega^{2}}{8\Delta}\cos(K\hat{x}),
\label{Hp_final}
\end{equation}
where $K=2k_{L}$.  In Eq.~(\ref{Hp_final}) we consider only the centre-of-mass motion of the atoms, as they are always assumed to be in the internal ground state.

\subsubsection{Time dependence of the laser intensity}

We consider the situation where the laser standing wave amplitude, turned on instantaneously at $t=0$, is switched on and off periodically.  Provided that the spectral width of each pulse is sufficiently narrow, i.e., the pulse duration $t_p \gg 1/ \Delta$, the time-dependent system can be described by Hamiltonian~(\ref{Hp_final}) with an additional time-dependent function $f(t)$ included in the standing-wave term.  The Hamiltonian is then
\begin{equation}
\hat{H}_{\delta} =\frac{ \hat{p}^{2}}{2M} - \frac{\hbar \Omega^{2}}{8\Delta}\cos(K\hat{x})f(t).
\label{H_pp_tdep}
\end{equation}
In the Raman-Nath regime, where the atomic centre-of-mass motion is negligible over the pulse duration, $f(t)$ can be approximated by a train of $\delta$-kicks \cite{Moore1995, Klappauf1999} \footnote{Finite pulse effects have been observed experimentally~\cite{Oskay2000}.  For simplicity, we do not consider these effects here as they play a relatively minor role, given an appropriate choice of parameters.}, i.e.,
\begin{equation}
f(t) \approx t_p \sum_{n=0}^{\infty}\delta (t-nT).
\label{ftfunc}
\end{equation}
In Eq.~(\ref{ftfunc}), $T$ is the time between successive pulses, and we scale by the pulse duration~$t_p$ such that $f(t)$ has unit amplitude in the limit $t_p \rightarrow 0$.

Finally, our model Hamiltonian for a thermal atomic gas subjected to an far-detuned pulsed standing-wave potential is~\footnote{More formal explorations of $\delta$-kicked rotor dynamics generally consider the sum  $\sum_{n=-\infty}^{\infty}\delta(t-nT)$ in the Hamiltonian~(\ref{H_final})~\cite{Reichl2004}.  However, when considering dynamics beginning at $t=0$, the time evolution generated by the Hamiltonian is independent of whether the lower limit of the sum is $n=-\infty$ or $n=0$.}
\begin{equation}
\hat{H}_{\delta} =\frac{ \hat{p}^{2}}{2M}  - \hbar
\phi_d \cos(K\hat{x})\sum_{n=0}^{\infty}\delta (t-nT),
\label{H_final}
\end{equation}
where we have defined an effective potential depth $\phi_{d}=\Omega^{2}t_{p}/8\Delta$.  Equation~(\ref{H_final}) is the Hamiltonian for a quantum $\delta$-kicked rotor~\cite{Reichl2004} (or, more correctly, a $\delta$-kicked particle, as the momentum spectrum is continuous rather than discrete).

\subsubsection{Spatial periodicity of the Hamiltonian \label{sec:spatial}}

Hamiltonian~(\ref{H_final}) is spatially periodic.  This implies conservation of the \textit{quasimomentum}, i.e., Bloch's theorem applies \cite{Kittel1996}.   To explain this, it is convenient to expand  the position and momentum operators into discrete and continuous components:
\begin{equation}
\begin{split}
\hat{x} &= K^{-1}( 2\pi \hat{l} +\hat{\theta}), \label{spatial_periodicity}\\
\hat{p} & =  \hbar K(\hat{k} + \hat{\beta}).
\end{split}
\end{equation}
The eigenvalues of $\hat{l}$ and $\hat{k}$ are integers, the eigenvalues of $\hat{\theta}\in [-\pi,\pi)$, and the quasimomentum operator~$\hat{\beta}$ is defined to take the eigenvalues of the continuous momentum component such that $\beta \in [-1/2,1/2)$.  Substituting the reformulated operators~(\ref{spatial_periodicity}) into Hamiltonian~(\ref{H_final}), yields
\begin{equation}
 \hat{H}_{\delta}=\frac{[\hbar K(\hat{k}+\hat{\beta})]^{2}}{2M}
 - \hbar\phi_{d}
\cos ( \hat{\theta} ) \sum_{n=0}^{\infty}\delta(t-nT).
\label{Hamiltonian_k_beta}
\end{equation}
The operator $\hat{\beta}$ commutes with both $\hat{\theta}$ and $\hat{k}$ \cite{Bach2005}, and hence with $\hat{H}_{\delta}$.  Thus, in the system of a $\delta$-kicked particle the laser field induces coupling only between momentum eigenstates differing in momenta by integer multiples of~$\hbar K$ \footnote{The momentum transfer $\hbar K$ corresponds physically to the momentum kick resulting from a stimulated two-photon interaction with the laser field, i.e., $\hbar K$ is equal to two photon recoils.}.

\subsection{Time evolution in the presence of quantum resonances and antiresonances \label{sec:time}}

\subsubsection{The Floquet operator}

The Floquet operator is the unitary time-evolution operator for one temporal period $T$ of the system.  The convention we have adopted defines the Floquet operator from just before one kick to just before the next.  The Floquet operator for the system governed by Hamiltonian~(\ref{Hamiltonian_k_beta}) is then \cite{Reichl2004}
\begin{equation}
 \hat{F}= \exp\left( -
\frac{i\hbar K^2 T}{2M }[\hat{k}+\hat\beta]^{2}
 \right) \exp ( i\phi_{d} \cos \hat{\theta}  ).
 \label{Floquet_k_beta}
\end{equation}
We consider $\hat{F}$ applied to a state with a particular value of the quasimomentum $\beta$, i.e.,
\begin{equation}
|\psi (\beta) \rangle = \sum_{k=-\infty}^{\infty}c_{k}|k+\beta \rangle.
\label{Eq:qmestate}
\end{equation}
Note that the dimensionless momentum eigenket $|k+\beta\rangle$ satisfies the eigenvalue equation
\begin{equation}
\hat{p}|k+\beta \rangle = \hbar K (k+\beta) |k+\beta \rangle,
\end{equation}
and the orthogonality condition
\begin{equation}
\langle k'+\beta' |k+\beta \rangle = \delta_{kk'} \delta(\beta-\beta').
\label{Eq:Orthog}
\end{equation}
As the quasimomentum is a conserved quantity (see Sec.~\ref{sec:spatial}), the operator $\hat{\beta}$ in the Floquet operator~(\ref{Floquet_k_beta}) can be replaced by its eigenvalue $\beta$ when acting on the quasimomentum eigenstate~(\ref{Eq:qmestate}), i.e.,
\begin{equation}
\hat{F}|\psi(\beta)\rangle
=
\hat{F}(\beta)
|\psi(\beta)\rangle,
\end{equation}
where
\begin{equation}
\hat{F}(\beta)=
\exp\left( -
\frac{i\hbar K^2 T}{2M}[\hat{k}+\beta]^{2}
 \right) \exp ( i\phi_{d} \cos \hat{\theta} ).
 \label{Eq:Floq_op}
\end{equation}

\subsubsection{Quantum resonances and antiresonances \label{sec:res_zero_qm}}

The time evolution described by the Floquet operator~(\ref{Eq:Floq_op}) can vary significantly depending on the system parameters.  We consider the strong driving regime, where the classical stochasticity parameter $\kappa \gtrsim 1.5$ \cite{Haake2001}.   In terms of parameters relevant to the quantum system described in this paper, $\kappa = K^{2}T\hbar\phi_{d}/M$.  In the case of strong driving dynamical localization is typically observed \cite{Grempel1982,Grempel1984,Fishman1993,Graham1992,Moore1994,Moore1995,Steck2000,Oskay2003,Bharucha1999,Ammann1998a,Vant2000,Doherty2000,dArcy2001b,Klappauf1999,Milner2000}, whereby  the atomic momentum spread increases linearly with time at the classical ``diffusion'' rate \cite{Lichtenberg1992}, i.e.,
\begin{equation}
\langle \hat{p}^{2}\rangle_{n}  = 
\left(
\frac{M}{KT}
\right)^{2} \frac{\kappa^{2}}{2}n
+\langle \hat{p}^{2}\rangle_{0}, 
\label{Eq:Class_Diff}
\end{equation}
up until a quantum break time \cite{Grempel1984,Moore1994,Moore1995}.  While dynamical localization is the more common scenario, the Floquet operator simplifies considerably for particular choices of the time $T$ between successive $\delta$-kicks.  Distinctive resonances and antiresonances then occur in the $\delta$-kicked particle system (e.g., \cite{Moore1995,Oskay2000,Duffy2004b,Ryu2006,Kanem2007}).

To illustrate resonance and antiresonance in an atom-optical context we consider their simplest manifestation, which occurs in the $\beta=0$ subspace; this means assuming the initial system state $| \psi (\beta=0) \rangle$ [see Eq.~(\ref{Eq:qmestate})].  Due to quasimomentum conservation, $\beta$ must always remain $=0$.  The accessible momentum spectrum is therefore discrete, consisting of integer multiples of $\hbar K$ only.  This can be mapped to the spectrum of an angular momentum, which consists of integer multiples of $\hbar$. The results presented here for the $\beta=0$ subspace are therefore equivalent to those gained from theoretical quantum studies of the ``genuine'' rotor~\cite{Casati1979,Reichl2004}.  

We consider the $\delta$-kicked particle for the case where the pulse periodicity $T$ is given by
\begin{equation}
T = \frac{\ell 2 \pi M}{\hbar K^2},
 \label{define_ell}
\end{equation}
where $\ell$ is a positive integer.  The Floquet operator~(\ref{Eq:Floq_op}), with $\beta=0$, is then $\hat{F}(0) = \exp( - i\ell \pi \hat{k}^{2}
 ) \exp (i\phi_{d} \cos \hat{\theta}).$  The free evolution operator $\exp(-i\ell \pi \hat{k}^2)$ is equivalent to $\exp(-i\ell \pi \hat{k})=\exp(i\ell \pi \hat{k}),$  because the eigenvalues of $\hat{k}$ are integer, and parity is conserved when an integer is squared.  The time evolution due to the Floquet operator $\hat{F}(0) = \exp( \pm i\ell \pi \hat{k}
 ) \exp (i\phi_{d} \cos \hat{\theta})$ is dramatically different depending on whether $\ell$ is even or odd.

For the case that $\ell$ is even, the free evolution operator collapses to the identity, i.e., $\hat{F}(0)= \exp(i\phi_{d} \cos \hat{\theta})$, and
\begin{equation}
\hat{F}^n(0)= e^{in\phi_{d} \cos \hat{\theta}}.
\label{beta_zero_resonance}
\end{equation}
Thus, after a time $nT$, the final system state is as if a single $\delta$-kick had been applied with a strength $n$ times the original kick strength~$\phi_d$.  This phenomenon is known as a \textit{quantum resonance}.  The shortest pulse periodicity for which resonance occurs, i.e., $T_T = 4 \pi M /\hbar K^2$ ($\ell=2$), is known as the Talbot time by analogy with the Talbot length in classical optics~\cite{Hecht2002}.

Considering now the general case where $\ell$ may be odd or even, we note that $[\hat{\theta},\hat{k}]=i$ \cite{Bach2005}, and therefore obtain
\begin{equation}
\begin{split}
\hat{F}^2(0)& =
e^{ -
i\ell \pi \hat{k}
 }e^{i\phi_{d} \cos \hat{\theta}}e^{
i\ell \pi \hat{k}
 }e^{i\phi_{d} \cos \hat{\theta}}, \\
 & = e^{i\phi_{d} \cos (\hat{\theta}-\ell \pi)} e^{i\phi_{d} \cos \hat{\theta}}, \\
 & = e^{i\phi_{d} [1+(-1)^\ell] \cos \hat{\theta}}.
 \end{split}
\end{equation}
If $\ell$ is even, we regain Eq.~(\ref{beta_zero_resonance}) for $n=2$. If $\ell$ is odd, $\hat{F}^2(0)=1$, and the system returns to its initial state every second pulse.  The recurrence of the initial state with period $2T$ is the phenomenon known as \textit{antiresonance}.  Higher-order antiresonances are also possible, where the initial state recurs after a larger number of iterations of the Floquet operator (see Sec.~\ref{sec:momentum})~\footnote{When the initial state of the system recurs with a period $NT$, for integer $\ell$, we refer to this as a higher-order antiresonance of order $N$.  We emphasise that the higher-order antiresonances described here are not to be confused with higher-order resonances that have been observed for pulse periodicities $T$ equal to $\ell T_T/ \jmath$ where $\ell$ and $\jmath$ are integers and $\jmath>2$~\cite{Kanem2007, Ryu2006}.}.

In this paper we consider the $\delta$-kicked particle for pulse periodicities $T=\ell T_T/2$, where $\ell$ is a positive integer [see Eq.~(\ref{define_ell})].  For the $\beta=0$ subspace, resonance occurs for even values of $\ell$, and antiresonance for odd values of $\ell$.  However, for a general $\beta$ subspace, the Floquet operator~(\ref{Eq:Floq_op}) is
\begin{equation}
\hat{F}(\beta) =
e^{ -i \pi
\beta^{2}\ell}e^{-i \pi \ell \hat{k}(1+ 2\beta)}
e^{ i\phi_{d} \cos \hat{\theta} }.
\label{Floquet_Talbot}
\end{equation}

The time evolution of the $\delta$-kicked particle, as governed by the Floquet operator~(\ref{Floquet_Talbot}), can be evaluated analytically.  In Secs.~\ref{sec:eevolve} and \ref{sec:evolve} we derive expressions for the system evolution, which encompass results presented previously by Bharucha {\it et al.}~\cite{Bharucha1999} for even $\ell$.

\subsubsection{Momentum eigenstate evolution \label{sec:eevolve}}

In this section we evaluate the time-evolution of a $\delta$-kicked particle, as described by the Floquet operator~(\ref{Floquet_Talbot}).  We consider a system initially prepared in the momentum eigenstate
\begin{equation}
|\Psi (t=0)\rangle = |k+\beta\rangle.
\label{init_state}
\end{equation}
Note that, once the evolution of state~(\ref{init_state}) is known, the evolution of any quasimomentum eigenstate [i.e., Eq.~(\ref{Eq:qmestate})], and in fact any general state, can also be determined.  A general treatment will be presented in Sec.~\ref{sec:evolve}.

After $n$ iterations of the Floquet operator, state $|\Psi (t=0)\rangle$ becomes
\begin{equation}
|\Psi (t=nT)\rangle = \hat{F}^n(\beta)|k+\beta \rangle.
\label{final_state}
\end{equation}
Due to quasimomentum conservation (see Sec.~\ref{sec:spatial}), Eq.~(\ref{final_state}) can be expanded using momentum eigenstates with the initial quasimomentum~$\beta$, i.e.,
\begin{equation}
|\Psi (t=nT)\rangle = \sum_{j=-\infty}^{\infty} c_{kj}(\beta, nT)|j+\beta\rangle,
\label{def_c}
\end{equation}
where the probability amplitudes $c_{kj}(\beta, nT)$ are given by
\begin{equation}
c_{kj} (\beta,nT) \delta(\beta-\beta ')= \langle j+\beta'| \hat{F}^n(\beta)|k+\beta\rangle.
\label{matrix_element_definition}
\end{equation}
We impose the normalisation condition $\sum_j |c_{kj}(\beta,nT)|^2=1$.   

The complex amplitudes $c_{kj}(\beta,nT)$ can be most simply evaluated using Eq.\ (\ref{matrix_element_definition}) and
\begin{equation}
\langle j+\beta'| \hat{F}^n(\beta)|k+\beta\rangle = \int dx \langle  j+\beta'| x \rangle \langle x |\hat{F}^n(\beta)| k+\beta\rangle,
\label{c_kj}
\end{equation}
where \begin{equation}
  \langle j+\beta'|x\rangle = \sqrt{\frac{K}{2\pi}} e^{ -i(j+\beta')Kx }.
 \label{plane_wave}
\end{equation}
A lengthy deduction (see Appendix~\ref{matrix_element}) determines that the matrix element $\langle x |\hat{F}^n(\beta) |k+\beta\rangle$ [see Eq.~(\ref{FzpApp})] is
\begin{equation}
\begin{split}
 \langle x | \hat{F}^n (\beta) |k+\beta \rangle = & \sqrt{\frac{K}{2\pi}} e^{
i 2n\beta \Upsilon} e^{ -i n\pi \beta^{2}\ell}  e^{  i(k+\beta)\left(Kx -2n \Upsilon \right)} \\
& \times \sum_{j=-\infty}^{\infty} e^{ijKx}J_{j}\left( \phi_d \frac{\sin (n \Upsilon)}{\sin \Upsilon} \right) i^j e^{-ij
(n+1)\Upsilon},
\label{Fzp}
\end{split}
\end{equation}
where, for convenience, we have defined the $\beta$-dependent variable
\begin{equation}
\Upsilon \equiv \frac{1}{2} \pi (1+2\beta)\ell.
\label{Upsilon}
\end{equation}
The matrix element~(\ref{c_kj}) then becomes
\begin{equation}
\begin{split}
\langle j+\beta'| \hat{F}^n(\beta)|k+\beta\rangle  = &  e^{ -i n\pi \beta^{2}\ell} e^{
-i 2 n k \Upsilon }
 \sum_{j'=-\infty}^{\infty} J_{j'}\left( \phi_d \frac{\sin (n \Upsilon)}{\sin \Upsilon} \right) \\
 & \times
 i^{j'} e^{-ij'  (n+1)\Upsilon}
 \frac{K}{2\pi}\int dx e^{i(k+j'-j+\beta-\beta')Kx}.
 \label{ckj_one}
 \end{split}
\end{equation}
Evaluating the Fourier integral in Eq.~(\ref{ckj_one}) [see Eq.~(\ref{Eq:Orthog})], and substituting the result back into Eq.~(\ref{matrix_element_definition}), we find that
\begin{equation}
 c_{kj}(\beta,nT) =  J_{j-k}\left( \phi_d \frac{\sin (n \Upsilon)}{\sin \Upsilon} \right)
 i^{j-k} e^{-i(j-k)  (n+1)\Upsilon} e^{-2in \Upsilon k}e^{ -i n\pi \beta^{2}\ell}.
 \label{coefficient}
\end{equation}

For the system initially prepared in the momentum eigenstate~(\ref{init_state}), the $q$th-order momentum moment after time $nT$ is 
\begin{equation}
\begin{split}
\langle \hat{p}^q\rangle_{n} & = (\hbar K)^{q} \sum_{j=-\infty}^{\infty} |c_{kj}(\beta,nT)|^2  (j+\beta)^{q} \\
 & = (\hbar K)^{q} \sum_{j=-\infty}^{\infty}
\left[
J_{j-k}\left( \phi_{d}
\frac{\sin(n \Upsilon)}{\sin\Upsilon}
\right)
\right]^2
(j+\beta)^{q}.
\label{mom_exp_es}
\end{split}
\end{equation}
The distinctiveness of resonant and antiresonant behaviour in the $\delta$-kicked particle can be concisely characterised by considering the time evolution of the second-order momentum moment [Eq.~(\ref{mom_exp_es}) with $q=2$], which is proportional to the kinetic energy.

\subsubsection{Evolution of an incoherent mixture \label{sec:evolve}}

The above treatment can be generalised to the case of an atom cloud at finite temperature, by considering an incoherent mixture of plane waves with an initial momentum distribution $D(\hbar K(k+\beta))=D_k(\beta)/\hbar K$.  The initial density operator for the system is taken to be
\begin{equation}
\begin{split}
\hat{\rho} &= \hbar K \int dp |p\rangle D(p) \langle p|,\\
&= \int_{-1/2}^{1/2} d\beta \sum_{k=-\infty}^{\infty}
|k+\beta\rangle
D_{k}(\beta)\langle k+\beta|,
\end{split}
\end{equation}
where we have assumed that the density matrix in the momentum representation is initially diagonal, i.e., there are no coherences in the system (as is the case for a thermal state).  After time $nT$, the density operator evolves to
\begin{equation}
\begin{split}
 \hat{\rho} = & \int_{-1/2}^{1/2} d\beta \sum_{k=-\infty}^{\infty}
\left[\sum_{j=-\infty}^{\infty}
c_{kj}(\beta,nT)|j+\beta\rangle
\right]
D_{k}(\beta) \\
& \times
\left[\sum_{j'=-\infty}^{\infty}
c_{kj'}^{*}(\beta,nT)
\langle j'+\beta|\right],
\end{split}
\label{final_density}
\end{equation}
where $c_{kj}(\beta,nT)$ is given by Eq.~(\ref{coefficient}).  Considering the diagonal elements of the density operator (\ref{final_density}), we find that the momentum distribution evolves to
\begin{equation}
D_k(\beta,t=nT)  = \sum_{j=-\infty}^{\infty}  |c_{jk}(\beta,nT)|^2
D_{j}(\beta),
\label{mom_dist}
\end{equation}
and the $q$th-order momentum moment is given by
\begin{equation}
\langle \hat{p}^q \rangle_{n}
 = (\hbar K)^q\int_{-1/2}^{1/2} d\beta
\sum_{j,k=-\infty}^{\infty}
\left|c_{kj}(\beta,nT) \right|^2
D_{k}(\beta)(j+\beta)^{q}.
\label{mom_mom_dist}
\end{equation}

\subsection{Numerical implementation \label{sec:numerical}}

\subsubsection{Monte-Carlo method}

The numerical results presented in this paper are generated using a Monte Carlo approach.  This is used in preference to a grid-based method, which could produce sampling errors if, for example, a disproportionate number of grid points were to coincide with strongly resonant or antiresonant values of the initial momentum.

The initial condition consists of $\cal{N}$ plane waves distributed according to the momentum distribution $D_k(\beta)$ (see Sec.~\ref{sec:evolve} and Sec.~\ref{sec:finiteT}).  The time evolution of each momentum eigenstate is evaluated by applying the Floquet operator~(\ref{Floquet_Talbot}), which is represented in a momentum basis~\cite{dArcy2001a}.  The final momentum distribution [see Eq.~(\ref{mom_dist})] is constructed by averaging the distributions resulting from each of the individual plane-wave time evolutions.

To illustrate our method we consider resonance and antiresonance for (i) the $\beta=0$ subspace (as described in Sec.~\ref{sec:res_zero_qm}), and (ii) a cold (but thermal) atomic gas~\cite{dArcy2001a}.  Sections~\ref{sec:num_beta0} and \ref{sec:finiteT} provide an introduction to the main content of this paper by summarising some relevant well known results~(e.g., \cite{Moore1995, Oskay2000}).

\subsubsection{Resonance and antiresonance in the $\beta=0$ subspace \label{sec:num_beta0}}

\begin{figure}[h!]
\includegraphics[width=8.5cm]{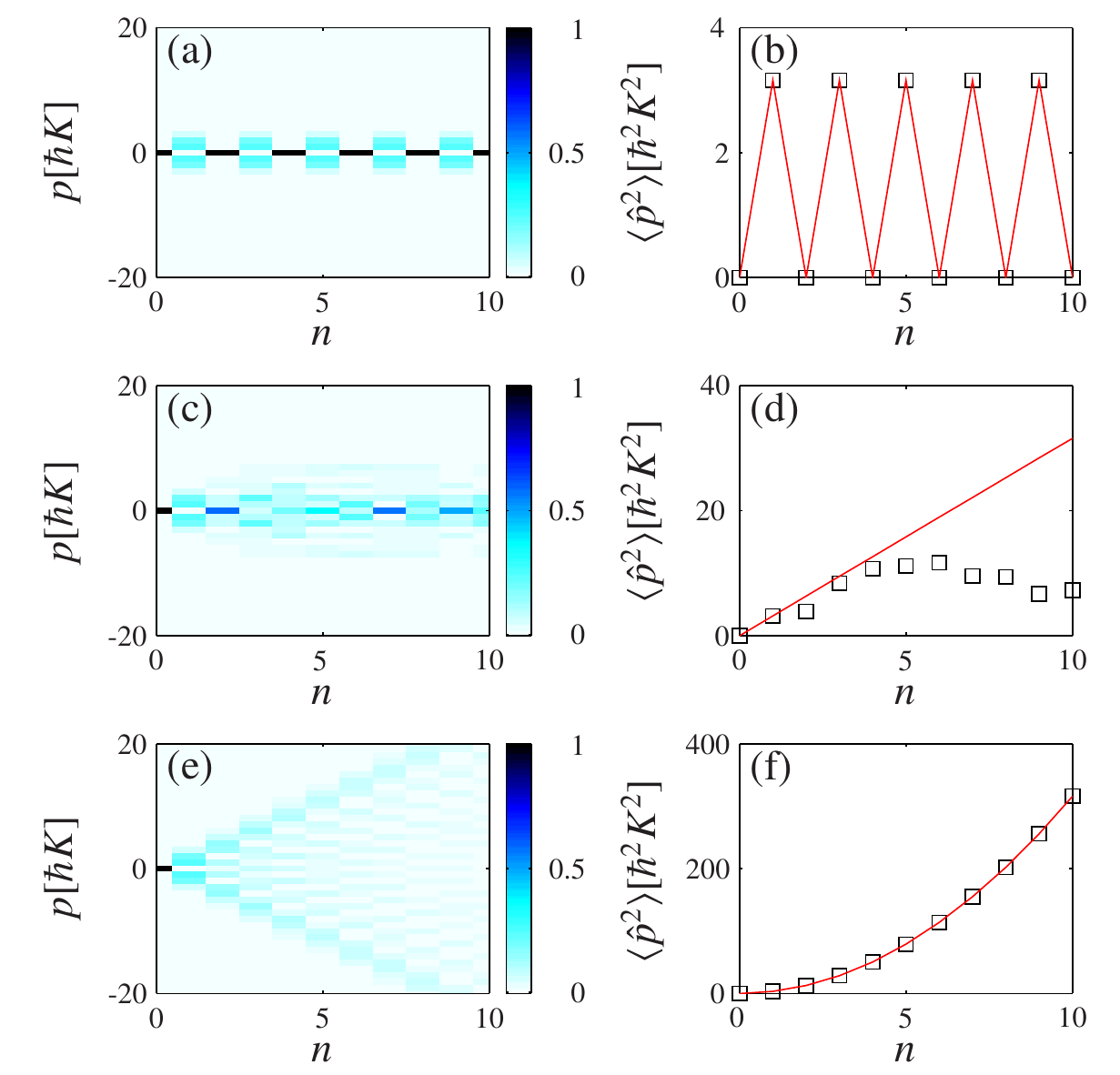}
\caption{Momentum distributions [(a), (c), and (e)] and corresponding second-order momentum moments [(b), (d), and (f)] of a $\delta$-kicked particle. In each case the initial condition is a zero-momentum eigenstate.  Parameters are ${\cal N}=1$, $\phi_d = 0.8\pi$, and (a), (b) $T = T_T/2$, (c), (d) $T = (1+\sqrt{5})T_T/2,$ and (e), (f) $T = T_T$.  The solid lines in (b), (d), and (f) are analytic results from Eqs.~(\ref{eqn:antires}), (\ref{Eq:Class_Diff}) (with $\langle \hat{p}^{2}\rangle_{0}=0$), and (\ref{eqn:res}), respectively.}
\label{fig:beta0}
\end{figure}

In Sec.~\ref{sec:res_zero_qm} we showed how the dynamics of the $\delta$-kicked particle depends dramatically on the pulse periodicity $T$.  In particular, resonance occurs in the $\beta=0$ subspace for even~$\ell$ (where $T= \ell T_T/2$ [see Eq.~(\ref{define_ell})]), and antiresonance is observed for odd values of~$\ell$.  Figure~\ref{fig:beta0} shows the atomic momentum distribution and the second-order momentum moment of a $\delta$-kicked particle, for the case where the system is initially in the zero-momentum eigenstate with unit amplitude.  We have chosen $\phi_{d}=0.8\pi$ as an illustrative value typical of recent experiments \cite{Oberthaler1999,Godun2000,dArcy2001a,dArcy2001b,dArcy2003,Schlunk2003a,Schlunk2003b,Ma2004,Buchleitner2006}.

Figures~\ref{fig:beta0}(a) and (b) are for the case $\ell = 1$.  We observe the characteristic feature of an antiresonance, i.e., recurrence of the initial state with period $2T$.  In contrast, Figs.~\ref{fig:beta0}(e) and (f) correspond to the resonant case $\ell=2$.  We observe quadratic growth in the second-order momentum moment, otherwise referred to as ballistic expansion~\cite{Oskay2000}.  Figures~\ref{fig:beta0}(c) and (d) show the system evolution for the case where $T$ is an irrational multiple of the Talbot time.  We observe dynamical localisation, characterised by classical-like diffusion until a quantum break time \cite{Grempel1984,Moore1994,Moore1995,Haake2001,Reichl2004}.

As described in Sec.~\ref{sec:res_zero_qm}, due to quasimomentum conservation 
the accessible momentum spectrum for the dynamics considered in Fig.~\ref{fig:beta0} can be mapped to the spectrum of an angular momentum.  The results presented in Fig.~\ref{fig:beta0} are therefore equivalent to those that would be obtained by studying the quantum $\delta$-kicked dynamics of a ``genuine'' rotor~\cite{Casati1979,Reichl2004}.  

\subsubsection{Resonance phenomena for a finite temperature cloud \label{sec:finiteT}}

In the case of a finite temperature cloud, all the quasimomentum subspaces are initially populated, not just the $\beta=0$ subspace.   Note that the Floquet operator governing the dynamics in each quasimomentum subspace is explicitly dependent on $\beta$ [see Eq.~(\ref{Eq:Floq_op})]. The overall dynamics can therefore exhibit some significant qualitative differences when compared to the case where the dynamics are restricted to the $\beta=0$ subspace (and, by extension, the dynamics of a ``genuine'' rotor).

We model a cold thermal cloud of atoms by considering a Gaussian initial momentum distribution  $D_k(\beta)$ with zero mean and standard deviation~$w$:
\begin{equation}
D_k(\beta) = \frac{1}{w \sqrt{2\pi}} \exp\left( \frac{-(k+\beta)^2}{2w^2}\right).
\label{Gauss}
\end{equation}
This corresponds to a Maxwell-Boltzmann distribution for free particles with temperature ${\cal T}_w=\hbar^{2}K^{2}w^{2}/mk_{B}$.  The second-order momentum moment of the kicked cloud, after time $nT$, is determined by substituting the momentum distribution~(\ref{Gauss}) into the general momentum moment evolution expression~(\ref{mom_mom_dist}) for $q=2$, to give
\begin{equation}
\begin{split}
\langle \hat{p}^2\rangle_{n}
 = &\frac{ \hbar^2 K^2}{w \sqrt{2\pi}}\int_{-1/2}^{1/2} d\beta
\sum_{j,k=-\infty}^{\infty}
\left[
J_{j-k}\left(
\phi_{d}
\frac{\sin(n\Upsilon)}{\sin\Upsilon}
\right)
\right]^{2} \\
& \times
e^{-(k+\beta)^{2}/2w^{2}}
(j+\beta)^{2}.
\end{split}
\label{mom_Gauss}
\end{equation}

\begin{figure}[h!]
\includegraphics[width=8.5cm]{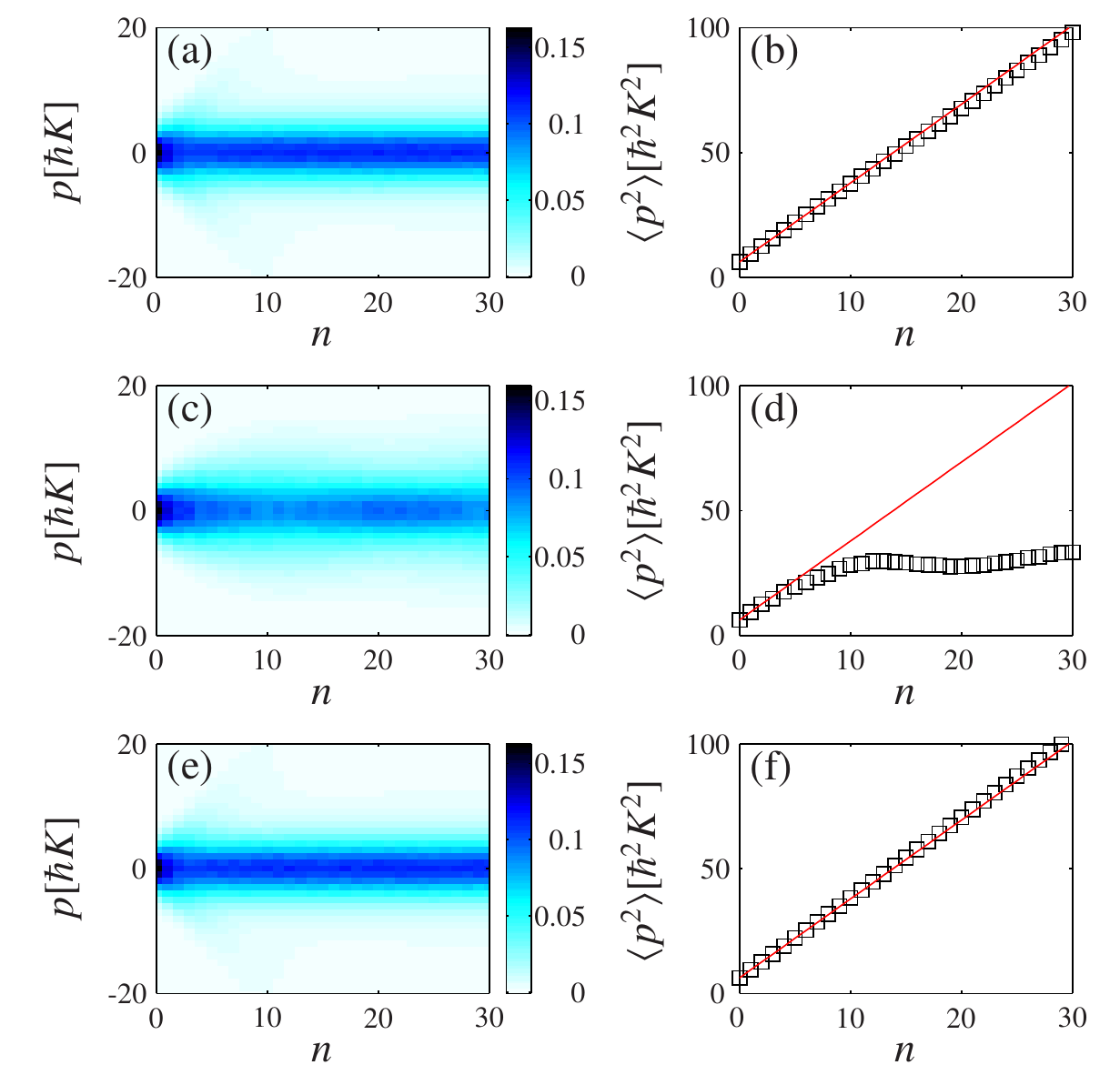}
\caption{Momentum distributions [(a), (c), and (e)] and corresponding second-order momentum moments [(b), (d), and (f)] of a $\delta$-kicked particle.  In each case the initial atomic momentum distribution is given by Eq.~(\ref{Gauss}).  Parameters are ${\cal N}=10000$, $\phi_d = 0.8\pi$, $w=2.5$, and (a), (b) $T = T_T/2$, (c), (d) $T = (1+\sqrt{5})T_T/2$, and (e), (f) $T = T_T$.  The solid lines in (b), (d), and (f) correspond to the linear  result~(\ref{Eq:Class_Diff}).}
\label{fig:finiteT}
\end{figure}

Figure~\ref{fig:finiteT} shows the atomic momentum distribution and the second-order momentum moment of a $\delta$-kicked cloud of noninteracting particles with initial standard deviation $w = 2.5$ (corresponding to approximately 5 $\mu$K in the case of caesium \cite{dArcy2001a}).  In contrast to the zero-momentum eigenstate case (see Fig.~\ref{fig:beta0}), the finite temperature cloud appears to behave identically for $\ell=1$ [Figs.~\ref{fig:finiteT}(a) and (b)] and $\ell=2$ [Figs.~\ref{fig:finiteT}(e) and (f)]~\footnote{Oskay {\it et al.}~\cite{Oskay2000} have shown that resonant and antiresonant dynamics can be distinguished at finite temperature because the trajectories of the ballistically expanding atoms are subtly different.  We consider the limit $t_p \rightarrow 0$ and, therefore, do not observe such differences.}.  For $\ell=1$ or 2, a small fraction of the cloud expands ballistically, but much of the distribution remains near zero momentum.  This clustering near the centre is due to antiresonant effects occuring concurrently with the quantum resonant dynamics. The result is limited ballistic expansion, which will be discussed in more detail in Sec.\ \ref{sec:resonance}. In Figs.~\ref{fig:finiteT}(c) and (d), where $T$ is an irrational multiple of the Talbot time, we observe dynamical localization rather than ballistic expansion.  The distribution in Fig.~\ref{fig:finiteT}(c), while truncated at the edges, appears flatter and broader compared to Figs.\ \ref{fig:finiteT}(a) and (e).

Figures~\ref{fig:finiteT}(b) and (f) show that the second-order momentum moment grows linearly.  This is dramatically different from either the periodic behaviour or the quadratic growth observed in Figs.~\ref{fig:beta0}(b) and (f), respectively.  Linear energy growth is characteristic of resonant phenomena in a thermal atomic cloud~\cite{Oskay2000}, and the slope of the line [see Eq.~(\ref{Eq:Class_Diff})] is well approximated by the classical diffusion rate~\cite{Moore1995,Lichtenberg1992}. Close inspection of Figs.~\ref{fig:finiteT}(b) and (f) reveals that the classical expression slightly overestimates the numerical data in the antiresonant case, and underestimates in the resonant case.  This will be clarified in Sec.~\ref{sec:temperature}.

\section{Momentum dependence of resonant and antiresonant features \label{sec:momentum}}

\subsection{Motivation}

In Secs.~\ref{sec:res_zero_qm} and \ref{sec:num_beta0} we considered the $\beta=0$ subspace, and found that resonance and antiresonance are clearly distinct.  In contrast, at finite temperature (see Sec.~\ref{sec:finiteT}), the system appears to behave identically irrespective of whether $\ell$ is even or odd ($T=\ell T_T/2$).  In the remainder of this paper we investigate in detail the transition between the behaviour observed in Fig.~\ref{fig:beta0} for a zero-momentum eigenstate, and that observed in Fig.~\ref{fig:finiteT} for a broad thermal distribution.  We consider only the cases where the pulse periodicity $T$ is a half-integer multiple of the Talbot time (i.e., $T=\ell T_T/2$).  In this section we begin by looking at the effect of the quasimomentum $\beta$ on resonant and antiresonant dynamics in the system.

\subsection{Resonance and antiresonance conditions \label{sec:res_analytic}}

We consider a system initially prepared in a momentum eigenstate [see Eq.~(\ref{init_state})].  A resonance in the system occurs when the free evolution operator collapses to the identity.  From the Floquet operator~(\ref{Floquet_Talbot}), ignoring the global $\beta$-dependent phase, we find that resonance occurs when
\begin{equation}
\frac{1}{2}(1+2\beta)\ell =m,
\label{res_cond}
\end{equation}
where $m$ is an integer.  In general, the probability that the initial state $|k+\beta \rangle$ is measured to have momentum $\hbar K (j+\beta)$, after a time $nT$, is  determined from Eq.~(\ref{coefficient}) to be \cite{Bharucha1999}
\begin{equation}
|c_{kj}(\beta,nT)|^2 = \left[
J_{j-k}\left( \phi_{d}
\frac{\sin(n \Upsilon)}{\sin\Upsilon}
\right)
\right]^2.
\label{prob}
\end{equation}
At resonance, when Eq.~(\ref{res_cond}) is satisfied, $\Upsilon = m \pi$ [see Eq.~(\ref{Upsilon})], so $\sin ( n\Upsilon) =0$ and $\sin\Upsilon=0$.  Taking the limit $\Upsilon\rightarrow m\pi$, Eq.~(\ref{prob}) becomes
\begin{equation}
|c_{kj}(\beta,nT)|^2 = [
J_{j-k} ( n \phi_{d} )]^2.
\end{equation}
As discussed in Sec.~\ref{sec:res_zero_qm}, the system responds as if a single $\delta$-kick had been applied with strength $n\phi_d$.

From Eq.~(\ref{prob}) we see that antiresonance occurs when
\begin{equation}
\frac{1}{2}(1+2\beta)\ell = m + \frac{1}{2},
\label{antires_2}
\end{equation}
for integer $m$.  In this case $\Upsilon = (m +1/2)\pi$, and the probability that the initial state is measured to have momentum $\hbar K (j+\beta)$ is
\begin{equation}
|c_{kj}(\beta,nT)|^2 = [
J_{j-k} ( \phi_d \sin(n\pi/2)
)
]^2.
\label{antires_analytic}
\end{equation}
For even values of $n$, $|c_{kj}(\beta,nT)|^2= \delta_{jk},$ and the initial state recurs.  Hence, every kick-induced diffraction is exactly inverted by a kick-induced contraction.

From Eq.\ (\ref{prob}) it is also apparent that higher-order antiresonances are possible.  An $N$th order antiresonance, for which the system returns to its initial state every $N$th kick, occurs when
\begin{equation}
\frac{1}{2}(1+2\beta)\ell =  \frac{m}{N},
\label{antires_high}
\end{equation}
where $m$ is a nonzero integer, and $m$ and $N$ have no common factors.  In this case,
\begin{equation}
|c_{kj}(\beta,nT)|^2 = \left[
J_{j-k}\left( \phi_{d}
\frac{\sin(n m \pi /N)}{\sin (m \pi /N)}
\right)
\right]^2.
\end{equation}
When $n$ is a multiple of the order $N$, $|c_{kj}(\beta,nT)|^2= \delta_{jk}$, and the system returns to its initial state \footnote{Note that when $N=2$, we regain the results of Eqs.~(\ref{antires_2}) and (\ref{antires_analytic})}.

\subsection{Role of the quasimomentum $\beta$ \label{sec:betascan}}

The $\beta$ dependence of the system evolution can be investigated in more detail by looking at the evolution of momentum eigenstates $|k+\beta \rangle$, where $k=0$ and $\beta \in [-1/2, 1/2).$  It is convenient to consider the momentum variance
\begin{equation}
\langle \langle \hat{p}^2 \rangle \rangle _n = \langle \hat{p}^2 \rangle_n - \langle \hat{p} \rangle_n^2.
\label{cumulant}
\end{equation}
The momentum variance~(\ref{cumulant}) displays the same signatures of resonant and antiresonant dynamics that have been described for the second-order momentum moment [see Figs.~\ref{fig:beta0}(b) and (f), and Sec.\ \ref{sec:eevolve}].  However, the variance is more useful in the respect that its evolution is independent of the discrete momentum $k$ [as we show in  Eq.~(\ref{cumu_indep_k}) below].  In Figs.~\ref{fig:beta0} and \ref{fig:finiteT},  the mean momentum $\langle \hat{p} \rangle_n$ is zero (or negligible) and it is not necessary to distinguish between the variance~(\ref{cumulant}) and the momentum moment $\langle \hat{p}^2 \rangle_{n}$.  However, in this section we consider the momentum variance~(\ref{cumulant}), because the mean momentum of the eigenstate $|k+\beta \rangle$ is in general nonzero.

The evolution of the momentum variance for a $\delta$-kicked particle initially in a momentum eigenstate $|k+\beta \rangle$ can be determined from Eq.~(\ref{mom_exp_es}).   Changing the summation index to $j'=j-k$, and using Eqs.~(\ref{eqnone}) and (\ref{eqntwo}), the evolution of the variance simplifies to
\begin{equation}
\langle \langle \hat{p}^2 \rangle \rangle _n = \hbar^2 K^2 \sum_{j'} \left[ J_{j'}  \left( \phi_{d}
\frac{\sin(n \Upsilon)}{\sin\Upsilon}
\right) \right]^2(j'+\beta)^2-\hbar^2 K^2 \beta^2.
\label{cumu_indep_k}
\end{equation}
The variance~(\ref{cumu_indep_k}) is independent of $k$ so we can explore the evolution of all momentum states $|p \rangle$ by considering only $k=0$ eigenstates where the quasimomentum $\beta$ spans the range $\beta \in [-1/2, 1/2).$

\begin{figure}[h!]
\includegraphics[width=8.5cm]{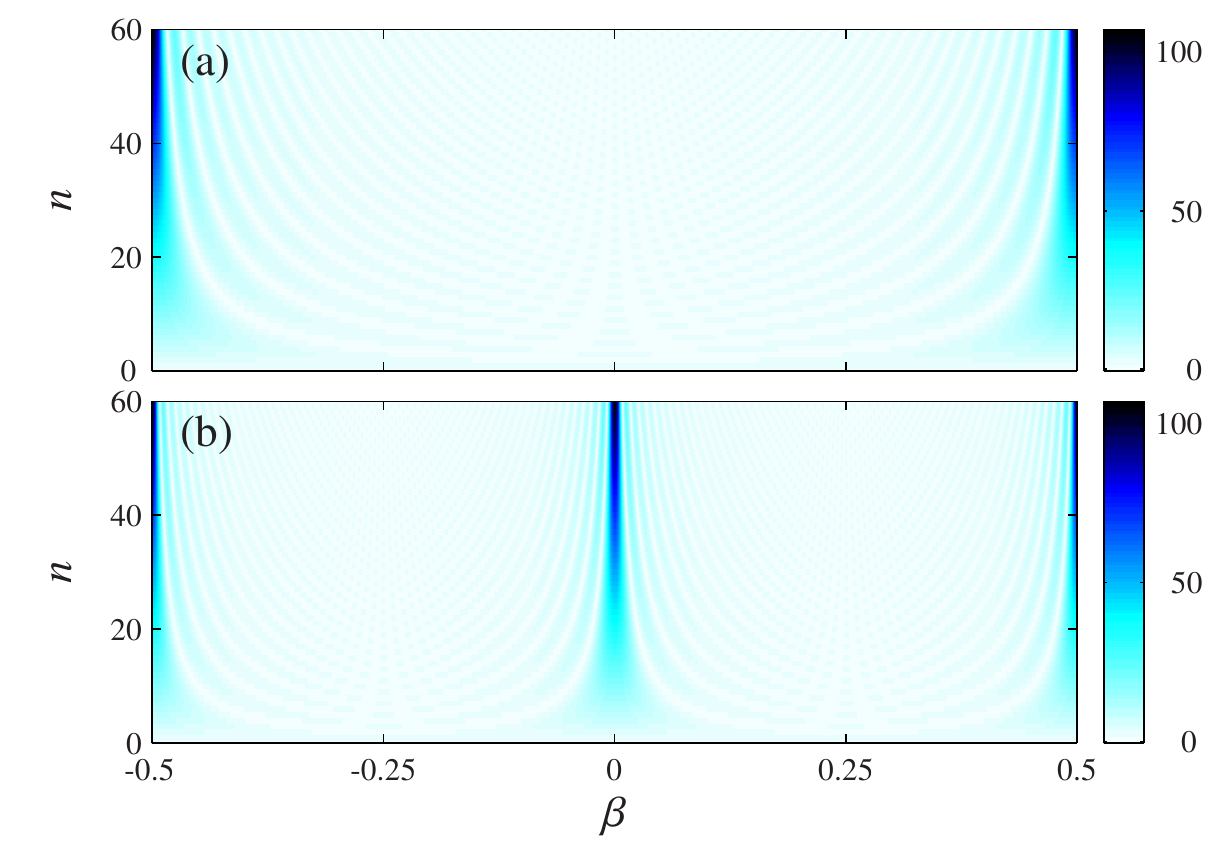}
\caption{$\langle \langle \hat{p}^2 \rangle \rangle_n ^{1/2}$ of a $\delta$-kicked particle initially in the plane-wave state $|\beta \rangle$.  Parameters are ${\cal N}=1$, $\phi_d = 0.8\pi$, and (a) $T = T_T/2$ and (b) $T=T_T$.}
\label{fig:betascan}
\end{figure}

Figure~\ref{fig:betascan} illustrates the $\beta$ dependence of the system dynamics.  In Fig.~\ref{fig:betascan}(a), for $\ell = 1$, an antiresonance occurs at $\beta=0$, as observed in Figs.~\ref{fig:beta0}(a) and (b).  Resonances occur at $\beta = \pm 1/2$.  In Fig.~\ref{fig:betascan}(b), for $\ell=2$, a resonance occurs at $\beta=0$, as observed in Figs.~\ref{fig:beta0}(e) and (f), and also at $\beta = \pm 1/2$.  Antiresonances are observed at $\beta = \pm 1/4$.  

The values of $\beta$ for which resonances and antiresonances occur can be determined by rearranging the resonance and antiresonance conditions, (\ref{res_cond}) and (\ref{antires_2}), respectively.  A resonance occurs in a particular $\beta$ subspace if $\beta = \beta^{\rm R}_m$, where $\beta^{\rm R}_m = m/\ell -1/2$ and $m$ is an integer.  The spacing between resonances is then $\beta^{\rm R}_{m+1}-\beta^{\rm R}_m = 1/\ell.$  An antiresonance occurs in a particular $\beta$ subspace if $\beta = \beta^{\rm A}_m$, where $\beta^{\rm A}_m = (2m+1)/2 \ell -1/2$.  Antiresonances occur exactly halfway between resonances, and are also separated by $\beta^{\rm A}_{m+1}-\beta^{\rm A}_m = 1/\ell.$

\subsection{Resonance and antiresonance width \label{sec:width}}

\subsubsection{Reconstruction loci of higher-order antiresonances}

\begin{figure}[h!]
\includegraphics[width=8.5cm]{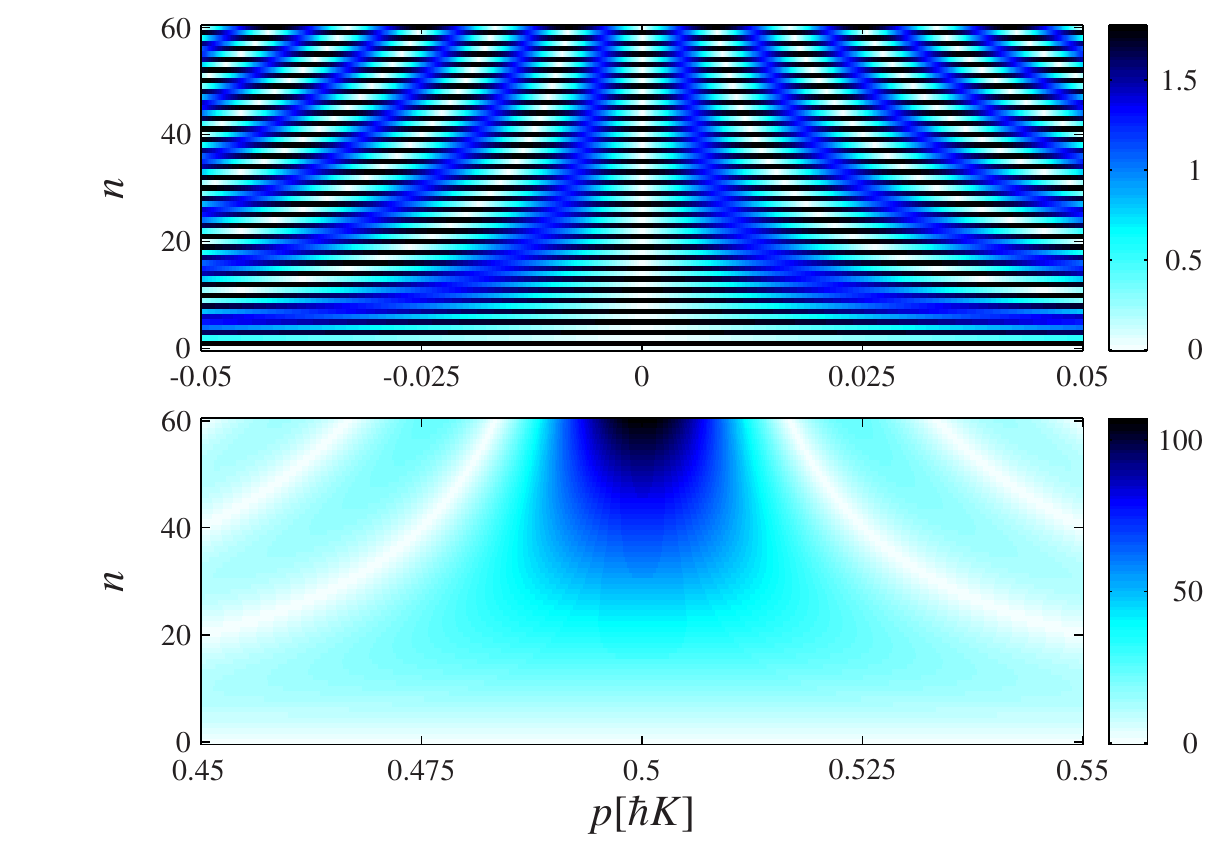}
\caption{$\langle \langle \hat{p}^2\rangle \rangle_n ^{1/2}$ of a $\delta$-kicked particle initially in the plane-wave state $|p \rangle$.  Parameters are ${\cal N}=1$, $\phi_d = 0.8\pi$, and $T = T_T/2$.}
\label{fig:betazooma}
\end{figure}

\begin{figure}[h!]
\includegraphics[width=8.5cm]{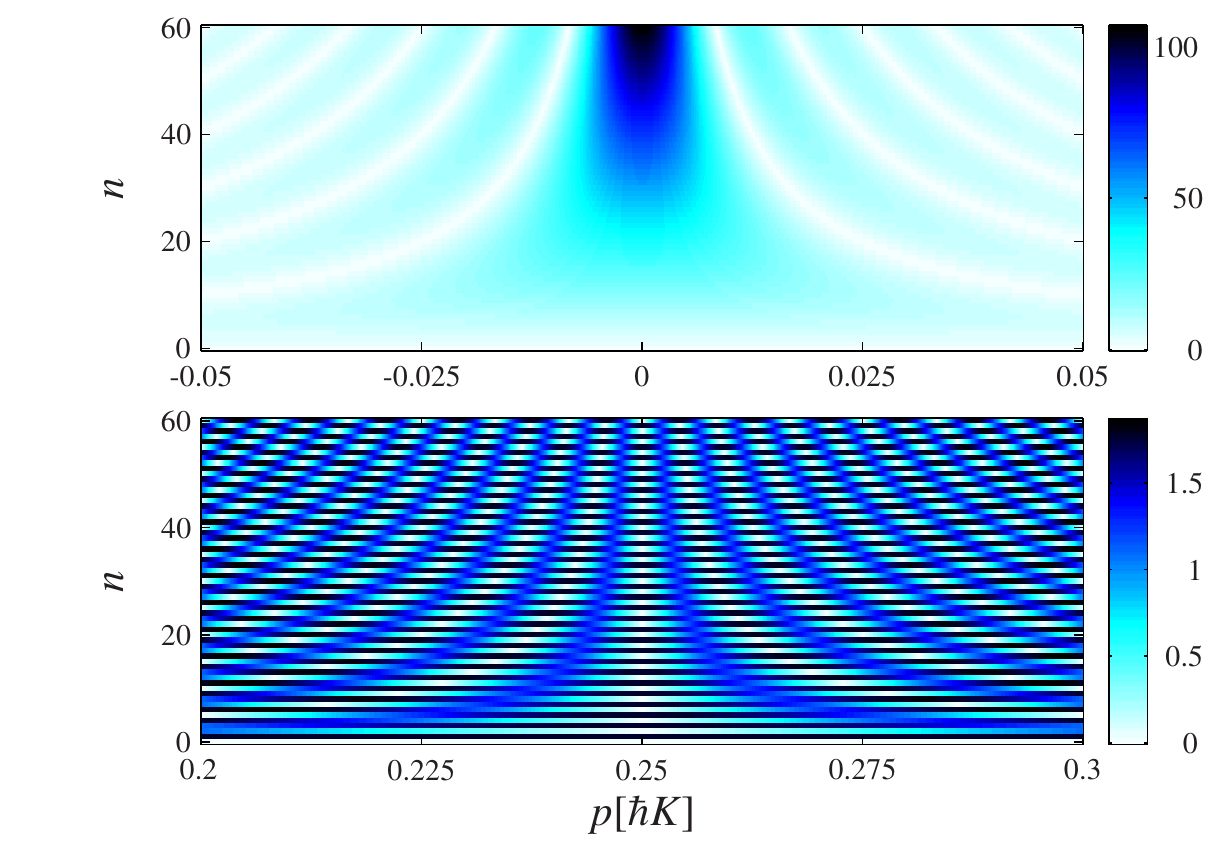}
\caption{$\langle \langle \hat{p}^2\rangle \rangle_n ^{1/2}$ of a $\delta$-kicked particle initially in the plane-wave state $|p \rangle$.  Parameters are ${\cal N}=1$, $\phi_d = 0.8\pi$, and $T = T_T$.}
\label{fig:betazoomb}
\end{figure}

The resonances and antiresonances in Fig.~\ref{fig:betascan} have a momentum width associated with them, which depends on $\ell$ and the kick number $n$.  As a precursor to deriving expressions for these widths, we zoom in on resonant and antiresonant features for $\ell = 1$ (see Fig.~\ref{fig:betazooma}) and $\ell=2$ (see Fig.~\ref{fig:betazoomb}).  In close proximity to the main resonant and antiresonant features, we observe hyperbolic curves of zero momentum variance. We call these \textit{reconstruction loci\/}, as they each correspond to a discontinuous series of points in time where higher-order antiresonances cause the initial condition to be periodically reconstructed.

\subsubsection{Resonance width}

Here we determine the locations of the higher-order antiresonances, at which the system returns to its initial state ($\langle \langle \hat{p}^2\rangle \rangle_n^{1/2}=0$).  We begin by rewriting Eq.~(\ref{antires_high}) as
\begin{equation}
\frac{1}{2}  (1+2\beta)\ell = m+m'/\bar{N},
\label{hoanti_new}
\end{equation}
 where $m$ and $m'$ are integers, and $0<|m'|<\lceil \bar{N}/2 \rceil$.  The integer $\bar{N}$ can be any kick number for which the initial state reconstructs as a result of an $N$th-order antiresonance, i.e., $\bar{N}$ is an integer multiple of the antiresonance order $N$.  A direct correspondence can now be made between the integer $m$ in Eq.~(\ref{hoanti_new}) and the integer $m$ in $\beta^{\rm R}_m$ [or, equivalently, in the resonance condition~(\ref{res_cond})].  In close proximity to a resonance, we evaluate the difference $\bar{\beta} = \beta - \beta^{\rm R}_m$ for which $\beta$ satisfies the higher-order antiresonance condition~(\ref{antires_high}).  We find that the terms involving $m$ cancel, and the equations $\bar{\beta} = m'/\bar{N} \ell$, where $0<|m'|<\lceil \bar{N} /2 \rceil$, describe the reconstruction loci in $\beta$ and $n$.

It is now possible to define a resonance width as follows.  The reconstruction loci closest to a resonance are described by $\bar{\beta} =  \pm 1/\bar{N} \ell$.  For $|m'|=1$, $\bar{N}$ and $N$ are equivalent, and the reconstruction loci can also be written as $\bar{\beta} =  \pm 1/ N \ell$.  The hyperbolic curves lying halfway between the resonance at $\beta^{\rm R}_m$ and the reconstruction loci $\bar{\beta} =  \pm 1/ N \ell$ we call \textit{transition loci}, as they are the sequence of points at which the wave-function begins the process of reconstruction.  The transition loci are described by $\bar{\beta} =  \pm 1/2 N \ell$.  We quantify the resonance width by the separation in $\beta$ between the two transition loci adjacent to the resonance, i.e.,
\begin{equation}
\delta \beta_R = \frac{1}{N \ell}.
\label{res_width}
\end{equation}

\subsubsection{Antiresonance width}

A measure of the second-order antiresonance width can be determined in a similar way to the resonance case, by considering the closest antiresonances (of any order) near an $N=2$ antiresonance.  In general, the higher-order antiresonances in close proximity to an $N=2$ antiresonance occur for $\bar{\beta}= \beta - \beta_m^A$, where $\beta$ satisfies the higher-order antiresonance condition~(\ref{antires_high}).  It is convenient to again use Eq~(\ref{hoanti_new}), but now $m'$ is constrained to the values $0<m'<\bar{N}$.  Making a direct correspondence between the integer $m$ in Eq.~(\ref{hoanti_new}) and the integer $m$ in $\beta^{\rm A}_m$, we find that $\bar{\beta} = (2m'-\bar{N})/2\bar{N} \ell.$  Defining $s=2m'-\bar{N}$ yields  $\bar{\beta} = s/2\bar{N} \ell$, where $s = (-|\bar{N}-2|, -|\bar{N}-2|+2, -|\bar{N}-2|+4, \dots, \bar{N}-2).$  

The reconstruction loci are described by $\bar{\beta} = s'/2\bar{N}\ell$, within the limits $0<|s'|<\bar{N}$, where $\bar{N}$ and $s'$ must be either both even, or both odd. Looking closely at the reconstruction loci adjacent to the antiresonances in Figs.~\ref{fig:betazooma} and \ref{fig:betazoomb}, it is possible to confirm that adjacent reconstruction loci correspond to $\bar{N}$ values of opposite parity \footnote{Due to the different colour scale, this level of detail is impossible to resolve in Fig.~\ref{fig:betascan}.}.  In particular, the $N=2$ antiresonance has even parity and the closest reconstruction loci, described by $\bar{\beta} = \pm 1/2 N \ell$ (taking $s'= \pm1$ and $N=\bar{N}$), have odd parity.

Transition loci can be defined which lie halfway between the $N=2$ antiresonance at $\beta^{\rm A}_m$ and the closest reconstruction loci described by $\bar{\beta} = \pm 1/2 N \ell$.  In the antiresonance case the transition loci $\bar{\beta} = \pm 1/4N\ell$ are the sequence of points for which the even parity of the $N=2$ antiresonance changes over to the odd parity of the closest higher-order antiresonances.  The width of a second-order antiresonance can be characterised by the separation in $\beta$ between the transition loci $\bar{\beta} = \pm 1/4N\ell$, i.e.,
\begin{equation}
\delta \beta_A = \frac{1}{2 N \ell}.
\label{antires_width}
\end{equation}
The antiresonance width~(\ref{antires_width}) is a factor of two narrower than the resonance width $\delta \beta_R$, as is evident from Figs.~\ref{fig:betazooma} and  \ref{fig:betazoomb}.

\section{Effect of the initial atom cloud temperature \label{sec:temperature}}

\subsection{Overview}

In Sec.~\ref{sec:momentum} we found that the evolution of the momentum variance for momentum eigenstates $|k+\beta \rangle$ depends dramatically on the quasimomentum $\beta$, but is independent of the discrete momentum $k$.  This means that the evolution of a $\delta$-kicked particle, with a Gaussian initial momentum distribution [see Eq.~(\ref{Gauss})], depends sensitively on the initial population of $\beta$ subspaces \cite{Wimberger2005a}.  In particular, if the initial momentum distribution is broad compared to the momentum separation $1/\ell$ between consecutive resonances (or antiresonances) [see Sec.~\ref{sec:betascan}], the whole range of individual momentum eigenstate evolutions is covered (from completely resonant to completely antiresonant).  In this case an averaged evolution results \cite{Bharucha1999,Oskay2000,Oskay2003}.

In this section we investigate in detail the effect of the initial momentum distribution width $w$ on the evolution of a $\delta$-kicked particle, or ensemble of particles.  We give analytic expressions for the second-order momentum moment evolution for both a zero-momentum eigenstate (see Fig.~\ref{fig:beta0}) and for a finite temperature cloud (see Fig.~\ref{fig:finiteT}).  We then use our knowledge of the $\beta$ dependence of momentum-eigenstate evolutions (described in Sec.~\ref{sec:momentum}), to understand the changeover between the zero-temperature regime where resonant and antiresonant features are distinct, and the finite temperature case where the system energy is observed to grow linearly [see Figs.~\ref{fig:finiteT}(b) and (f)].

Note that for the remainder of this paper we consider atomic momentum distributions symmetric around zero.  Thus, it is equivalent to consider either the second-order momentum moment or the momentum variance [see Eq.~(\ref{cumulant})].  We consider the former for notational convenience.

\subsection{Effect of temperature on resonant dynamics \label{sec:resonance}}

For a zero-momentum eigenstate, resonance occurs for even $\ell$ ($T=\ell T_T/2$).  At a resonance, the second-order momentum moment evolves quadratically as
\begin{equation}
\langle \hat{p}^2 \rangle_{n} = \hbar^2 K^2 \frac{1}{2} \phi_d^2 n^2,
\label{eqn:res}
\end{equation} 
where we have taken Eq.~(\ref{momentum_k0}) with $\beta=0$ and $\ell$ even.  The quadratic dependence of the second-order momentum moment with the kick number $n$ is illustrated in Fig~\ref{fig:beta0}(f).  In contrast, the energy of a finite temperature cloud grows linearly at a resonance, as shown in Fig.~\ref{fig:finiteT}(f).  In the latter case, the growth rate is well approximated by the classical diffusion result [given by Eq.~(\ref{Eq:Class_Diff})].

\begin{figure}[h!]
\includegraphics[width=8.5cm]{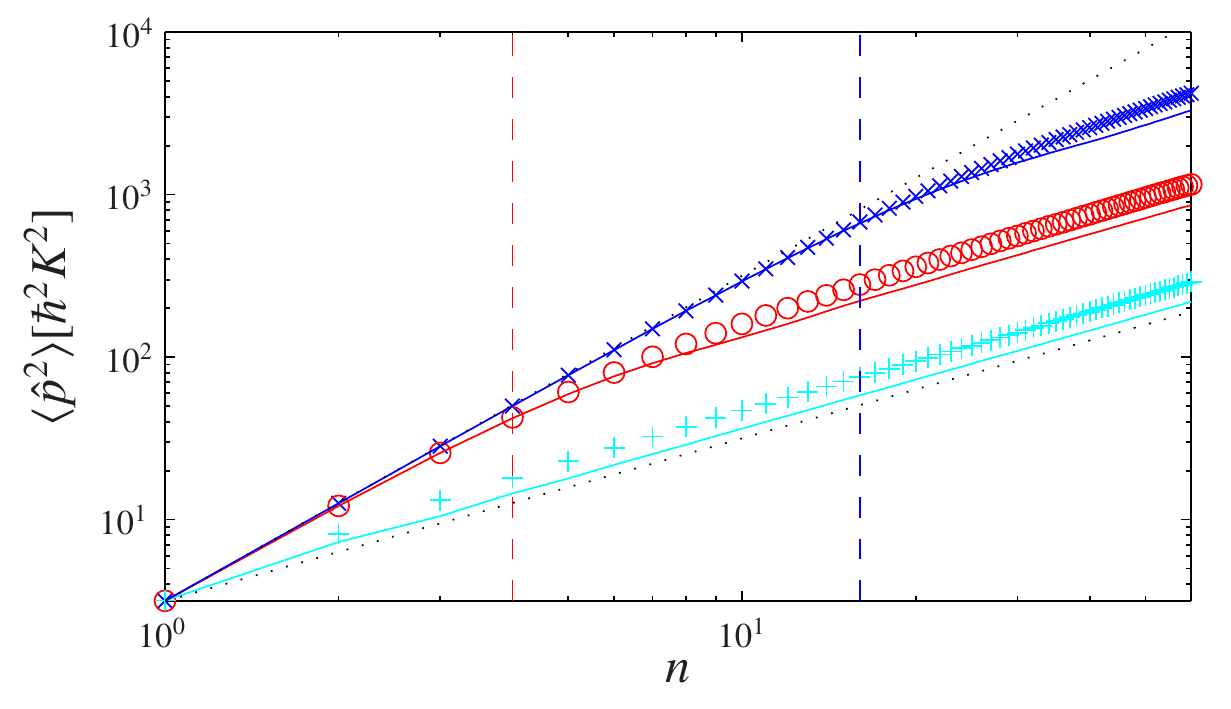}
\caption{Second-order momentum moment of a $\delta$-kicked atom cloud.  The initial state is a Gaussian momentum distribution with standard deviation ({\color{blue} $\times$})~$w=1/128$, ({\color{red} $\circ$})~$w=1/32$, and  ({\color{cyan} $+$})~$w=1/8$.  Parameters are ${\cal N}=10000$, $\phi_d = 0.8\pi$, and $T = T_T$.  The solid lines are analytic results for square distributions [see Eq.~(\ref{sq_result})], chosen with an initial standard deviation in agreement with the corresponding Gaussian case.  The upper dotted line corresponds to the quadratic growth~(\ref{eqn:res}), and the lower dotted line corresponds to the classical linear growth~(\ref{Eq:Class_Diff}), taking $\langle p^2 \rangle _0=0$.  The vertical dashed lines indicate $n=n_R$ [see Eq.~(\ref{threshold_time})].}
\label{fig:res}
\end{figure}

Figure~\ref{fig:res} (symbols) shows the evolution of the second-order momentum moment for three initial Gaussian momentum distributions with different values of the standard deviation $w$.  We observe that the momentum moment initially follows the quadratic growth of the zero-momentum eigenstate case (upper dotted line in Fig.~\ref{fig:res}) but, for a finite temperature cloud, the ballistic expansion becomes limited at a particular kick number, denoted $n_R$.  Above $n_R$ the rate of energy growth begins to revert to the classical diffusion result~(\ref{Eq:Class_Diff}) (indicated by the lower dotted line in Fig.~\ref{fig:res}).  For large $w$, the energy growth is bounded below by the classical result~(\ref{Eq:Class_Diff}) [see Fig.~\ref{fig:finiteT}(f)].

The behaviour observed in Fig.~\ref{fig:res} can be understood by considering the range of $\beta$ subspaces involved in the system evolution (see Sec.~\ref{sec:momentum}).  The resonance width was found in Sec~\ref{sec:width} to be $\delta \beta_R = 1/N\ell$ [see Eq.~(\ref{res_width})].  Provided that the initial momentum width of the cloud is well within the resonance width, resonant dynamics dominate the system evolution.  Conversely, if the initial momentum width is larger than the resonance width, higher-order antiresonances play a significant role and the ballistic expansion is limited.  

Quantitatively, we define a kick number $n_R$ to indicate the number of kicks at which the ballistic expansion becomes limited.  We define $n_R$ by requiring that two standard deviations of the initial Gaussian momentum distribution lie within the resonance width $\delta \beta_{R}$, i.e.,
\begin{equation}
n_{R} \equiv \frac{1}{4w \ell}.
\label{threshold_time}
\end{equation}
Equation~(\ref{threshold_time}) ensures that $95.4$~\% of the atom cloud  expands ballistically up until the kick number $n_R$ (indicated by the vertical dashed lines in Fig.~\ref{fig:res}).

To derive an analytic expression for the second-order momentum moment evolution of a finite temperature cloud, we consider a $\delta$-kicked particle with a square initial momentum distribution (see Appendix~\ref{sec:linear_growth}).  We find that the second-order momentum moment evolves as
\begin{equation}
\langle \hat{p}^2 \rangle_n = \hbar^2 K^2 \left[ \frac{\epsilon^2}{3} + \frac{1}{2} \phi_d^2 n 
+ \frac{\phi^2_d}{2 \epsilon \pi \ell} \sum_{m=1}^{n-1} \frac{n-m}{m}
 \sin (2m\pi \ell \epsilon )\right],
\label{sq_result}
\end{equation}
where $\epsilon$ is the half-width of the initial momentum distribution, and $m$ is integer [see Eq.~(\ref{sq_evolve}) with even $\ell$].  If the boundaries of the square distribution lie exactly on resonant or antiresonant values of the quasimomentum (i.e., $\epsilon = m'/2\ell$, for positive integer $m'$ within the limits $0<m' \leq \ell$), all the $\beta$ subspaces are populated equally (see Sec.~\ref{sec:betascan} and  Fig.~\ref{fig:betascan}).  Equation~(\ref{sq_result}) then simplifies to
\begin{equation}
\langle \hat{p}^2 \rangle_n = \hbar^2 K^2 \left( \frac{1}{2} \phi_d^2 n   + \frac{\epsilon^2}{3} \right).
\label{eqn_sq_half}
\end{equation}
Noting that $\langle \hat{p} \rangle ^2_0=\hbar^2 K^2 \epsilon^2 / 3$, we recover the linear growth in the momentum moment evolution, as predicted by the classical result~(\ref{Eq:Class_Diff}).  

The moment evolution~(\ref{sq_result}) of the square distribution is in good agreement with the moment evolution of a Gaussian momentum distribution with the same initial standard deviation, as shown in Fig.~\ref{fig:res}.  The moment evolution of a square distribution slightly underestimates the moment evolution for the corresponding Gaussian distribution, because the Gaussian has a relatively larger proportion of the initial population in close proximity to the resonance ($p=0$).

\subsection{Effect of temperature on antiresonant dynamics}

For a zero-momentum eigenstate, antiresonance occurs for odd integers $\ell$.  At the antiresonance, $|c_{0j}(0)|^2=J_j^2(\phi_d)$ for odd $n$, and $|c_{0j}(0)|^2=\delta_{j0}$ for even $n$ [see Eq.~(\ref{antires_analytic})].  The second-order momentum moment evolves as 
\begin{equation}
\begin{split}
\langle \hat{p}^2 \rangle _n & = \hbar^2 K^2 \frac{1}{2} \phi_d^2 \sin^2(n\pi/2),\\
& = \hbar^2 K^2 \frac{1}{4} \phi_d^2 [1-(-1)^n],
\label{eqn:antires}
\end{split}
\end{equation}
where we have taken Eq.~(\ref{momentum_k0}) with $\beta=0$ and $\ell$ odd.  The oscillation of the second-order momentum moment with period $2T$ is illustrated in Fig~\ref{fig:beta0}(b).  

\begin{figure}[h!]
\includegraphics[width=8.5cm]{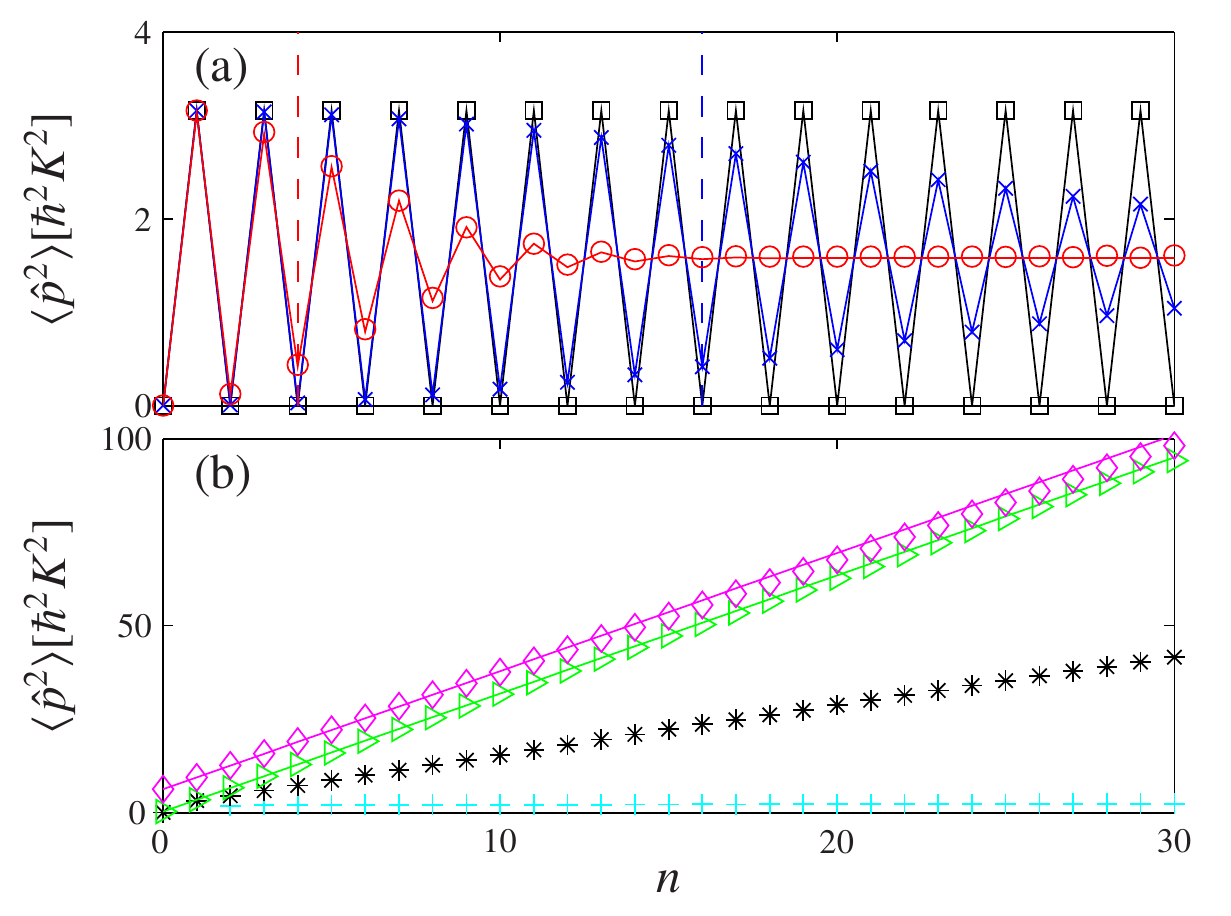}
\caption{Second-order momentum moment of a $\delta$-kicked atom cloud.  The initial state is a Gaussian momentum distribution with standard deviation (a) ({\color{black} $\square$}) $w=0$, ({\color{blue} $\times$}) $w=1/128$, and  ({\color{red} $\circ$}) $w=1/32$, and (b) ({\color{cyan} $+$}) $w=1/8$, ({\color{black} $*$}) $w=1/4$, ({\color{green} $\triangleright$}) $w=1/2$, and ({\color{magenta} $\diamond$}) $w=2.5$.  Parameters are ${\cal N}= 10000$, $\phi_d = 0.8\pi$, and $T = T_T/2$.  In (a) the solid lines are given by Eq.~(\ref{analytic_antires}) and the vertical dashed lines indicate $n=n_A$ [see Eq.~(\ref{nA})].  The solid lines in (b) are given by Eq.~(\ref{Eq:Class_Diff}).}
\label{fig:antires}
\end{figure}

The effect of the initial atomic momentum width on the second-order momentum moment evolution at an antiresonance is shown in Fig.~\ref{fig:antires}.  The evolution can be separated into two regimes: (i) when $w \ll 1/2\ell$ periodic oscillations are observed [see  Fig.~\ref{fig:antires}(a)], and (ii) for $w  \gtrsim 1/2\ell$ the momentum moment grows linearly [see Fig.~\ref{fig:antires}(b)].  For large $w$, the energy growth is bounded above by the classical result~(\ref{Eq:Class_Diff}) [see Fig.~\ref{fig:finiteT}(b)].

As in the resonance case, we can define a kick number $n_A$ below which the majority of plane-wave components in the distribution antiresonate with period $2T$.  Requiring that two standard deviations of the initial Gaussian atomic momentum distribution lie within the antiresonance width $\delta \beta_A$ [see Eq.~(\ref{antires_width})], we find that
\begin{equation}
n_A = \frac{1}{8 w \ell}.
\label{nA}
\end{equation}
In Fig.~\ref{fig:antires}(a), the kick numbers $n_A$ are indicated by the vertical dashed lines.  For kick numbers less than $n_A$ the initial state recurrence with period $2T$ is clearly observed.

In the low temperature regime [where $w \ll 1/2\ell$, as in Fig.~\ref{fig:antires}(a)], we observe oscillations with period $2T$ even after $n=n_A$.  However, the oscillations decay to a constant value.  The momentum moment evolution in the low temperature regime is well approximated by 
\begin{equation}
\langle \hat{p}^2 \rangle_n
\approx  \frac{1}{4}\hbar^2 K^2 \phi_d^2 \left[ 1 - \cos (n\pi)  e^{-2n^2 \pi^2 \ell^2 w^2} \right]+\hbar^2 K^2 w^2,
\label{analytic_antires}
\end{equation}
as shown in Fig.~\ref{fig:antires}(a).  The derivation of Eq.~(\ref{analytic_antires}) is presented in Appendix~\ref{sec:antires_der}.

The observed decay in the periodic oscillations occurs when the initial momentum width of the cloud is comparable to, or larger than, the antiresonance width $\delta \beta_A$.  Provided that $w \ll 1/2\ell$, higher-order antiresonances begin to play a role in the system dynamics, but resonances do not.  In this regime the evolution of the different $\beta$ subspaces can be described as an oscillation of the same amplitude as the $\beta=0$ subspace, but with a different phase.  This means that the different $\beta$ subspaces gradually become out of phase with each other, leading to a dephasing in the oscillation of the second-order momentum moment.  The decay of the oscillations has been observed experimentally by Duffy {\it et al.}~\cite{Duffy2004b}.  

For the case where $w  \gtrsim 1/2\ell$ [see Fig.~\ref{fig:antires}(b)], certain initially populated $\beta$ subspaces fulfill the resonance condition~(\ref{res_cond}), so the second-order momentum moment is observed to grow linearly.  For sufficiently broad initial momentum distributions, the slope of the linear moment growth tends to the classical diffusion rate [see Eq.~(\ref{Eq:Class_Diff})].

\section{Discussion and Conclusions}

In this paper we have investigated theoretically the quantum resonant and antiresonant dynamics of $\delta$-kicked cold atoms.  In particular we have investigated how the behaviour changes as one considers progressively broader initial momentum distributions. That is, as one considers increasing the initial temperature of the atomic cloud from the ideal case of zero Kelvin.  We have identified transition times, as a function of the initial momentum width, after which the quantum resonant and antiresonant dynamics clearly deviate from the ideal zero temperature case.  This can be understood by examining the time-evolution of individual momentum eigenstates, which varies greatly with the quasimomentum.  Very cold temperatures are typically achieved with an atomic Bose-Einstein condensate, which introduces a mean field.  For the results discussed by Duffy \textit{et al.}~\cite{Duffy2004b}, mean field effects are not very significant compared to the effect of the sample's initial momentum width. The analysis presented in this paper can then still be applied.  It should be noted, however, that with increased strength of the interatomic interactions, this may no longer hold \cite{Zhang2004,Liu2006}.

Typical temperatures achieved with an initially magneto-optically trapped sample, such as used in the Oxford experiment~\cite{Oberthaler1999,Godun2000,dArcy2001a,dArcy2001b,dArcy2003,Schlunk2003a,Schlunk2003b,Ma2004,Buchleitner2006} (temperature  $\sim5$ $\mu$K, corresponding to a Gaussian momentum distribution with $w \sim 2.5 \hbar K$) observe a linear-with-time increase of the mean kinetic energy.  In this case there is very little to distinguish resonant from antiresonant evolution \cite{Bharucha1999,Oskay2000, dArcy2001a,Sadgrove2004}.  Narrower initial momentum distributions are achievable, however.  For example: the Rubidium 87 Bose-Einstein condensed sample of Duffy \textit{et al.}~\cite{Duffy2004b} had an initial width of $w\sim 0.043 \hbar K$ (with $\lambda=4\pi/K=780$nm; 1D Raman-cooled caesium with $w\sim0.062 \hbar K$ \cite{Reichel1995}, and Bose-Einstein condensed caesium with  $\sim 0.008 \hbar K$ \cite{Kraemer2004} ($\lambda=895$nm) is achievable; and samples of 1D Raman-cooled sodium with $w\sim0.0016 \hbar K$ \cite{Kasevich1991} ($\lambda=589$nm) have been reported.  Experimental exploration of the regimes described in this paper is therefore quite possible.

\section*{Acknowledgements}
We thank the UK EPSRC (Grant  no.\ EP/D032970/1), and  PLH thanks Durham University, for support.  We also thank C. S. Adams, S. L. Cornish, I. G. Hughes, and M. P. A. Jones, for useful discussions.

\begin{appendix}

\section{Evaluation of the Floquet operator matrix element \label{matrix_element} }

\subsection{Application of one Floquet operator}

In this Appendix we calculate the matrix element $\langle x |\hat{F}^n(\beta)|k+\beta\rangle$.  We begin by considering the effect of applying the Floquet operator $\hat{F}(\beta)$ to the momentum eigenket $|k+\beta\rangle$.  The simplest approach is to use that
\begin{equation}
\hat{F}(\beta)|k+\beta \rangle = \int dx
\hat{F}(\beta)|x\rangle \langle x |k+\beta \rangle.
\label{Floquet_momentum_ket}
\end{equation}
Substituting the Floquet operator~(\ref{Floquet_Talbot}) into Eq.~(\ref{Floquet_momentum_ket}) yields
\begin{equation} \label{floquet2}
 \hat{F}(\beta)|k+\beta\rangle
= e^{ -i \pi
\beta^{2}\ell} \int dx e^{ -i \hat{k} K \gamma }e^{ i\phi_{d} \cos(K\hat{x}) } |x\rangle \langle x |k+\beta\rangle ,
\end{equation}
where we have defined
\begin{equation} \label{Kfn}
 K \gamma \equiv \pi  (1+2\beta)\ell .
\end{equation}
The kicking term in the Floquet operator introduces a spatially dependent phase,
\begin{equation}
e^{ i\phi_{d}\cos(K\hat{x})} |x\rangle=|x\rangle e^{i\phi_{d}\cos(Kx)},
\label{op_law}
\end{equation}
and the free evolution leads to a position kick~\footnote{The translation operator $\exp(-i \gamma \hat{p}/\hbar)$ transforms a position eigenket as $\exp (-i \gamma \hat{p}/\hbar )|x \rangle = |x+\gamma \rangle.$  Substituting for the discrete and continuous momentum components [see Eq.~(\ref{spatial_periodicity})], and replacing the quasimomentum operator $\hat{\beta}$ by its eigenvalue $\beta$, we have that $\exp (-i \gamma \hat{k} K ) \exp (-i \gamma \beta K) |x \rangle = |x+\gamma \rangle$.}, i.e.,
\begin{equation}
 e^{
-i \hat{k} K \gamma }|x\rangle = e^{
i \beta K \gamma } |x+\gamma \rangle.
\label{floquet5}
\end{equation}
The matrix element $\langle x |k+\beta\rangle$ in Eq.~(\ref{floquet5}) is known [see Eq.~(\ref{plane_wave})], so we find that
\begin{equation}
\label{floquet6}
 \hat{F}(\beta)|k+\beta\rangle = \sqrt{\frac{K}{2\pi}} e^{
i \beta K \gamma } e^{ -i \pi \beta^{2}\ell}
   \int dx
|x+\gamma \rangle e^{i(k+\beta)Kx} e^{i\phi_{d}\cos(Kx)}.
\end{equation}

\subsection{Application of subsequent Floquet operators}

The time evolution of state~(\ref{floquet6}) due to a second $\delta$-kick, and free evolution for time $T=\ell T_T/2$, can be found in a similar way, to be
\begin{equation}
\begin{split}
 \hat{F}^2(\beta)|k+\beta \rangle
= & \sqrt{\frac{K}{2\pi}} e^{
i 2\beta K \gamma } e^{ -i 2\pi \beta^{2}\ell}  \int dx |x+2\gamma \rangle  \\
& \times e^{i(k+\beta)Kx}e^{ i\phi_{d} [ \cos(Kx)+\cos(Kx+K\gamma) ]}.
\end{split}
\end{equation}
Subsequently, the effect of the full time-evolution operator $F^n(\beta)$ is
\begin{equation}
\begin{split}
 \hat{F}^n(\beta)|k+\beta \rangle = & \sqrt{\frac{K}{2\pi}} e^{
i n\beta K \gamma } e^{ -i n\pi \beta^{2}\ell} \int dx | x +n\gamma \rangle e^{ i(k+\beta)Kx} \\
 & \times e^{ i\phi_{d} \left[ \cos(Kx)+\sum_{j=1}^{n-1}\cos\left( K
x+jK\gamma\right) \right] }.
\label{iteration}
\end{split}
\end{equation}

\subsection{Spatial representation}

The final state~(\ref{iteration}) can be projected onto the ket $|x'\rangle$, and the orthogonality relation $\langle x | x' \rangle = \delta(x-x')$ means that the integral can be evaluated.  That yields
\begin{equation}
\begin{split}
 \langle x |\hat{F}^n(\beta) |k+\beta \rangle = & \sqrt{\frac{K}{2\pi}} e^{
i n\beta K \gamma } e^{ -i n\pi \beta^{2}\ell} e^{ i(k+\beta)K(x-n\gamma)}  \\
& \times e^{i\phi_{d} \left[ \cos ( Kx-Kn\gamma ) + \sum_{j=1}^{n-1}\cos ( Kx-nK\gamma+jK\gamma  ) \right]}. \label{difficult_sums}
\end{split}
\end{equation}
Equation~(\ref{difficult_sums}) can be simplified by defining the variable
\begin{equation}
q_j  \equiv (n-j)K \gamma,
 \label{q_j}
\end{equation}
to give
\begin{equation}
\begin{split}
\langle x |\hat{F}^n(\beta) |k+\beta \rangle = &\sqrt{\frac{K}{2\pi}} e^{
i n\beta K \gamma } e^{ -i n\pi \beta^{2}\ell} e^{i(k+\beta) (Kx-q_0)} \\
& \times e^
{i\phi_{d} \sum_{j=0}^{n-1}\cos (Kx-q_j ) }.
\label{eqn_half_way}
\end{split}
\end{equation}
Defining $\xi \equiv \sum_{j=0}^{n-1}
\cos q_{j}$ and $\zeta  \equiv \sum_{j=0}^{n-1} \sin q_{j}$, Eq.~(\ref{eqn_half_way}) can be further simplified to
\begin{equation}
\begin{split}
 \langle x | \hat{F}^n(\beta)  |k+\beta \rangle = & \sqrt{\frac{K}{2\pi}} e^{
i n\beta K \gamma } e^{ -i n\pi \beta^{2}\ell}  e^{  i(k+\beta)\left(Kx-q_0\right) } \\
& \times e^{i\phi_{d} \xi\cos(Kx)
+i\phi_{d}\zeta  \sin(Kx)}.
\label{manipulate}
\end{split}
\end{equation}

\subsection{Bessel function expansion}

Invoking the Bessel function expansions
\begin{equation}
\begin{split}
e^{ i \phi_d \xi \cos(Kx) }
=& \sum_{j=-\infty}^{\infty}
i^j J_j(\phi_d \xi)
e^{ij Kx} ,
\\ e^{ i \phi_d \zeta \sin(Kx) }
=& \sum_{j=-\infty}^{\infty}
J_j(\phi_d \zeta) e^{ij Kx},
\label{comb2}
\end{split}
\end{equation}
we can rewrite Eq.~(\ref{manipulate}) as
\begin{equation}
\begin{split}
 \langle x |\hat{F}^n(\beta) |k+\beta \rangle = & \sqrt{\frac{K}{2\pi}} e^{
i n\beta K \gamma } e^{ -i n\pi \beta^{2}\ell}  e^{ i(k+\beta)\left(Kx-q_0\right)} \sum_{j=-\infty}^{\infty} e^{ijKx} \\
& \times \sum_{j'=-\infty}^{\infty}
e^{ij'\pi/2}
J_{j'}(\phi_d \xi) J_{j+j'}(\phi_d \zeta).
\label{manipulate2}
\end{split}
\end{equation}
Using Graf's addition theorem \cite{Abramowitz1964}, Eq.~(\ref{manipulate2}) becomes
\begin{equation}
\begin{split}
\langle x | \hat{F}^n(\beta) |k+\beta \rangle = &\sqrt{\frac{K}{2\pi}}  e^{
i n\beta K \gamma } e^{ -i n\pi \beta^{2}\ell}  e^{  i(k+\beta)\left(Kx-q_0\right) } \\
& \times \sum_{j=-\infty}^{\infty} e^{ijKx}J_{j}(\omega)e^{ij\chi},
\label{Exponential_term}
\end{split}
\end{equation}
where  $\omega$ and $\chi$ are real and defined by $\omega e^{i\chi} \equiv \phi_d ( i\xi+\zeta ).$  It is useful to define the function
\begin{equation}
\mu \equiv \xi+i\zeta = \sum_{j=0}^{n-1} e^{iq_j},
\label{mu_eqn}
\end{equation}
from which $\xi$ and $\zeta$ can be deduced via $\xi = (\mu +\mu^*)/2$ and $\zeta = (\mu - \mu^*)/2i$.  Substituting $q_j$ [from Eq.~(\ref{q_j})] into Eq.~(\ref{mu_eqn}), we find that
\begin{equation}
\mu = e^{inK\gamma}
\sum_{j=0}^{n-1} e^{-i j K \gamma },
\label{mu_simple}
\end{equation}
and evaluating the geometric sum gives
\begin{equation}
\mu = e^{
in  K \gamma}
\left[
\frac{1-e^{- in K \gamma
}}{1-e^{- i K \gamma
}}
\right].
\end{equation}
Using standard trigonometric identities, we find that
\begin{equation}
\begin{split}
\xi = & \sin(n \Upsilon)[
\cos(n \Upsilon)\cot\Upsilon - \sin(n \Upsilon)
], \\
\zeta = & \sin(n \Upsilon)[
\cos(n \Upsilon)+\cot\Upsilon\sin(n \Upsilon)
],
\end{split}
\end{equation}
where we have defined the parameter $\Upsilon = K \gamma /2$ [this is consistent with the definition of Eq.~(\ref{Upsilon})].  Finally,
\begin{equation}
\frac{\omega e^{i\chi}}{\phi_{d}}=\zeta + i\xi =
\frac{\sin(n \Upsilon)}{\sin\Upsilon}ie^{-i(n+1)\Upsilon},
\label{omega_result}
\end{equation}
and substituing back into Eq.~(\ref{Exponential_term}), the matrix element $\langle x | \hat{F}^n (\beta) |k+\beta \rangle $ is determined to be
\begin{equation}
\begin{split}
 \langle x | \hat{F}^n (\beta) |k+\beta \rangle = & \sqrt{\frac{K}{2\pi}} e^{
i 2n\beta \Upsilon} e^{ -i n\pi \beta^{2}\ell}  e^{  i(k+\beta)\left(Kx -2n \Upsilon \right)} \\
& \times \sum_{j=-\infty}^{\infty} e^{ijKx}J_{j}\left( \phi_d \frac{\sin (n \Upsilon)}{\sin \Upsilon} \right) i^j e^{-ij  (n+1)\Upsilon}.
\label{FzpApp}
\end{split}
\end{equation}

\section{Momentum-moment evolution for a low temperature atom cloud}

\subsection{Second-order momentum moment evolution for a momentum eigenstate with $k=0$ \label{sec:beta_small}}

We consider a system that is initially prepared in the momentum eigenstate $|\beta\rangle$, i.e., $k=0$ and $\beta \in [-1/2, 1/2).$  After a time $nT$, the second-order moment is given by [see Eq.~(\ref{mom_exp_es}) with $q=2$ and $k=0$]
\begin{equation}
\begin{split}
\langle \hat{p} ^2 \rangle_n
& = \hbar^2 K^2 \sum_{j=-\infty}^{\infty}
\left[ J_{j}\left( \phi_{d} \frac{\sin (n \Upsilon)}{\sin\Upsilon}
\right) \right]^2 (j+\beta)^2,
\label{psq}
\end{split}
\end{equation}
where $\Upsilon$ is defined in Eq.~(\ref{Upsilon}).  For compactness, we define the $\beta$ dependent variable
\begin{equation}
\eta = \phi_{d} \frac{\sin (n \Upsilon) }{\sin \Upsilon },
\end{equation}
and then Eq.~(\ref{psq}) can be expanded to give
\begin{equation}
\langle \hat{p}^2 \rangle_n
 = \hbar^2 K^2\sum_{j=-\infty}^{\infty}
 J_{j}^2(\eta)
 [j^2+2j\beta +\beta^2].
\label{KE_moment}
\end{equation}

To evaluate Eq.~(\ref{KE_moment}), we consider each the three terms on the right-hand side, in reverse order.  To evaluate the third term we use the identities~\cite{Abramowitz1964}
\begin{equation}
J_{-j}(\eta) = (-1)^j J_j (\eta),
\label{oddJ}
\end{equation}
and
\begin{equation}
J_0^2(\eta)+2\sum_{j=1}^{\infty}J_j^2(\eta) = 1,
\label{sumJs}
\end{equation}
to find that
\begin{equation}
\sum_{j=-\infty}^{\infty}
 J_{j}^2(\eta) \beta^2 = \beta^2.
 \label{eqnone}
\end{equation}
The second term can be evaluated, using Eq.~(\ref{oddJ}), to give
\begin{equation}
\sum_{j=-\infty}^{\infty}
 J_{j}^2(\eta)  2j\beta  =0.
 \label{eqntwo}
\end{equation}
To evaluate the first term in Eq.~(\ref{KE_moment}) we use the identity~\cite{Abramowitz1964} 
\begin{equation}
J_{j+1}(\eta)+J_{j-1}(\eta) = \frac{2j}{\eta}J_j(\eta),
\end{equation}
and Eq.~(\ref{eqnone}), to find that
\begin{equation}
\begin{split}
\sum_{j=-\infty}^{\infty}
j^2 J_{j}^2(\eta) = & \frac{1}{4} \eta^2 \sum_{j=-\infty}^{\infty} \left[ J_{j+1} (\eta) +J_{j-1}(\eta)\right]^2, \\
  = &\frac{1}{2} \eta^2\left[ 1+ \sum_{j=-\infty}^{\infty} J_{j+1} (\eta) J_{j-1} (\eta)\right].
\label{term_3}
\end{split}
\end{equation}
Using the addition theorem~\cite{Abramowitz1964}
\begin{equation}
J_j(\eta+\eta') = \sum_{j'=-\infty}^{\infty} J_{j'}(\eta) J_{j-j'}(\eta'),
\end{equation}
along with the identity $J_{j}(-\eta) = (-1)^jJ_j(\eta)$, and Eq.~(\ref{oddJ}), Eq.~(\ref{term_3}) becomes
\begin{equation}
\sum_{j=-\infty}^{\infty}
j^2 J_{j}^2(\eta) =  \frac{1}{2}\eta^2 \left[ 1+J_2(0) \right] =  \frac{1}{2} \eta^2.
\end{equation}
Therefore, the second-order momentum moment evolves as
\begin{equation}
\begin{split}
\langle \hat{p} ^2 \rangle_n
& = \hbar^2 K^2 \left( \frac{\phi_d^2}{2} \frac{\sin^2 (n\Upsilon)}{\sin^2\Upsilon} +\beta^2 \right).
\label{momentum_k0}
\end{split}
\end{equation}

\subsection{Second-order momentum moment evolution for a square distribution \label{sec:linear_growth}}

We consider a square initial atomic momentum distrubition of the form
\begin{equation}
D_k(\beta) = \left\{ \begin{array}{cc}  \frac{1}{2 \epsilon} \delta_{k0}, &  |\beta|<\epsilon\\
0, &  |\beta|>\epsilon. \end{array} \right.
\label{square}
\end{equation}
The initial standard deviation is $\epsilon/\sqrt{3}$, and the second-order momentum moment evolves as 
\begin{equation}
\langle \hat{p}^2 \rangle _n = \frac{\hbar^2 K^2}{2\epsilon} \int_{-\epsilon}^{\epsilon} d\beta \sum_{j=-\infty}^{\infty} J^2_{j}(\eta) [j + \beta]^2, \label{eqn1a}
\end{equation}
where we have taken Eq.~(\ref{mom_mom_dist}) with $q=2$ and $k=0$, and with  $D_k(\beta)$ given by Eq.~(\ref{square}).  The sum over $j$ in Eq.~(\ref{eqn1a}) can be evaluated as shown in Sec.~\ref{sec:beta_small}, to yield
\begin{equation}
\begin{split}
\langle \hat{p}^2 \rangle _n & = \frac{\hbar^2 K^2}{2\epsilon} \int_{-\epsilon}^{\epsilon}d\beta \left(\frac{1}{2} \eta^2 +\beta^2 \right), \\
& = \hbar^2 K^2 \left(\frac{\epsilon^2}{3}+\int_{-\epsilon}^{\epsilon} d\beta \frac{\phi_d^2}{4\epsilon} \frac{\sin^2(n\Upsilon)}{\sin^2\Upsilon} \right).
\end{split}
\label{psq_one}
\end{equation}
To evaluate the remaining integral, we make a change of variables such that
\begin{equation}
\langle \hat{p}^2 \rangle _n = \hbar^2 K^2 \left(\frac{\epsilon^2}{3}+ \frac{\phi_d^2}{4\epsilon \pi \ell} \int_{\pi \ell (1-2\epsilon)/2}^{\pi \ell (1+2\epsilon)/2}\frac{\sin^2(n\Upsilon)}{\sin^2\Upsilon} d\Upsilon  \right).
\end{equation}
Then using the expansion~\cite{Halkyard2007}
\begin{equation}
\frac{\sin^2(n\Upsilon)}{\sin^2 \Upsilon} = n+2 \sum_{m=1}^{n-1} (n-m) \cos 2m \Upsilon,
\end{equation}
where  $m$ is integer, we find that
\begin{equation}
\begin{split}
\langle \hat{p}^2 \rangle_n = & \hbar^2 K^2 \Biggl( \frac{\epsilon^2}{3} + \frac{1}{2} \phi_d^2 n 
\\&
+ \frac{\phi^2_d}{2 \epsilon \pi \ell} \sum_{m=1}^{n-1} \frac{n-m}{m} (-1)^{m \ell}
\sin (2 m\pi \ell \epsilon)\Biggr).
\label{sq_evolve}
\end{split}
\end{equation}

\subsection{Second-order momentum moment evolution for a Gaussian distribution \label{sec:antires_der}}

Here we consider the evolution of the second-order momentum moment for the case where the initial momentum distribution is Gaussian (see Sec.~\ref{sec:evolve}).  In particular, we consider an ultracold gas for which the momentum width of the initial distribution is very small compared to the two-photon recoil momentum $\hbar K$, i.e., $w \ll 1$.  Thus, initially the majority of atoms have $k=0$, and their $\beta$ values are typically much less than the maximum possible value $|\beta|=1/2$.  The second-order momentum moment evolution is described by Eq.~(\ref{mom_Gauss}) with $k=0$:
\begin{equation}
\langle \hat{p}^2 \rangle_n
= \frac{\hbar^2 K^2}{w \sqrt{2\pi}}\int_{-1/2}^{1/2} d\beta
\sum_{j=-\infty}^{\infty}
J_{j}^2(\eta)
e^{-\beta^{2}/2w^{2}}
(j+\beta)^{2}.
\label{moment_one}
\end{equation}
In the regime where $w \ll 1$, we can approximate Eq.~(\ref{moment_one}) by extending the limits of the $\beta$ integral to infinity, since the additional contribution to the integral will be negligible.  Evaluating the sum over $j$ as shown in Sec.~\ref{sec:beta_small}, yields
\begin{equation}
\begin{split}
\langle \hat{p} ^2 \rangle_n
& \approx \frac{\hbar^2 K^2}{w \sqrt{2\pi}}\int_{-\infty}^{\infty} d\beta \left(\frac{1}{2} \eta^2 +\beta^2\right) e^{-\beta^{2}/2w^{2}}
, \\
& = \hbar^2 K^2w^2+\frac{\hbar^2 K^2 \phi_d^2}{2w\sqrt{2\pi}}\int_{-\infty}^{\infty} d\beta \frac{\sin^2 (n\Upsilon) }{\sin^2\Upsilon} e^{-\beta^{2}/2w^{2}}.
\label{mom_exp}
\end{split}
\end{equation}

In the case that an antiresonance occurs at $\beta=0$, i.e., $\ell$ is odd, we can write $\Upsilon = \pi( m +1/2 + \beta \ell)$, where $m$ is an integer [see Eq.~(\ref{antires_2})].  Thus, $\sin \Upsilon = (-1)^m \cos (\pi\beta \ell)$ and $\sin (n\Upsilon) = (-1)^{nm} \sin (n\pi/2+n\pi\beta \ell)$.  In the regime where $|\beta \ell| \ll 1/2,$ we can write that $\sin \Upsilon \approx (-1)^m,$ which leads to the considerable simplification that
\begin{equation}
\begin{split}
\left\langle \left(\frac{\hat{p}}{\hbar K}\right)^2\right\rangle_n
& \approx w^2+\frac{\phi_d^2}{2w\sqrt{2\pi}}\int_{-\infty}^{\infty} d\beta \sin^2 (n\Upsilon)
e^{-\beta^{2}/2w^{2}}, \\
& = w^2 +\frac{\phi_d^2}{4} \left[ 1 -\frac{1}{ w\sqrt{2\pi}} \int_{-\infty}^{\infty} d\beta \cos (2
n\Upsilon) e^{-\beta^{2}/2w^{2}} \right].
\label{psq_an}
\end{split}
\end{equation}
It is reasonable to take $|\beta \ell| \ll 1/2$ since $w \ll 1$ implies that the only nonzero contributions to the momentum moment evolution are from small $\beta$.  
  
Using that $\Upsilon = \pi( m +1/2 + \beta \ell)$, we write $\cos (2 n \Upsilon) = \cos (n\pi) \cos (2 n \pi \beta \ell),$ and then the integral in Eq.~(\ref{psq_an}) can be evaluated as follows:
\begin{equation}
\begin{split}
\left\langle \left(\frac{\hat{p}}{\hbar K}\right)^2\right\rangle_n
  &\approx  w^2 +\frac{\phi_d^2}{4} \left[ 1-\frac{\cos (n\pi)}{w\sqrt{2\pi}}  \int_{-\infty}^{\infty} d\beta \cos (2 n \pi \beta \ell) e^{-\beta^{2}/2w^{2}} \right],\\
&  =w^2 +\frac{\phi_d^2}{4} \left[ 1 - \frac{\cos (n\pi)}{w\sqrt{2 \pi}} \int_{-\infty}^{\infty} d\beta e^{i 2n \pi \beta \ell} e^{-\beta^{2}/2w^{2}} \right], \\
& =  w^2 +\frac{\phi_d^2}{4} \left[ 1 - \cos (n\pi)  e^{-2n^2 \pi^2 \ell^2 w^2} \right].
\end{split}
\label{analytic_antires_app}
\end{equation}

\end{appendix}


\end{document}